# Strong men create good memes: Far-right memes as means of authoritarian legitimation in Russia

**Polina Smirnova**[1,*] · **Mykola Makhortykh**[2]
[1] Department of Social and Political Sciences, University of Milan, Milan, Italy
[2] Institute of Communication and Media Studies, University of Bern, Bern, Switzerland
*Corresponding author: polina.smirnova@unimi.it
ORCID: 0009-0005-6667-6884 (P. Smirnova), 0000-0001-7143-5317 (M. Makhortykh)

## Abstract

This study examines how Russian far-right communities on Telegram shape perceptions of political figures through meme-based visual narratives. In Russia, such groups are central because they articulate the ideological foundations of the war against Ukraine and reflect the regime's drift toward ultranationalist rhetoric. Drawing on a dataset of 404,013 images from expert-selected far-right Telegram channels, the study uses automated face recognition, CLIP-based image embeddings, and HDBSCAN clustering with UMAP to identify memes featuring Russian (Putin, Medvedev) and foreign politicians (Zelensky, Biden, Trump) and to uncover recurrent visual patterns in their representation. Findings show that far-right memes co-produce state propaganda by generating new bottom-up narratives of legitimation and delegitimation that situationally align with state ideology. By framing authoritarian leaders as heroic and their opponents as corrupt, weak, or non-human, far-right actors act as informal co-creators of authoritarian legitimacy within Russia's informational autocracy.

**Keywords:** authoritarian legitimacy; memes; Telegram; far-right online communities; Russia

The concept of informational autocracy has become central to contemporary research on modern authoritarianism. Guriev and Treisman (2019) define informational autocracies as regimes that maintain power primarily through information manipulation, as mass repression became increasingly costly, less effective, and more visible to the outside world. Recent studies expand this framework, showing how authoritarian regimes cultivate legitimacy through public narratives, ideological messaging, and other mechanisms of autocratic legitimation (Dukalskis & Gerschewski, 2017; Rosenfeld & Wallace, 2024).

Digital technologies play a crucial role in information autocracies, sustaining regime stability and legitimation. Frantz et al. (2020) show that dictatorships increasingly rely on digital repression to reduce the risk of protest. Platforms such as Telegram are often used to disseminate propaganda, creating closed networks that facilitate communication among regime supporters (Oleinik, 2024) and engaging external audiences through bots and trolls (Khazhomia, 2025).

Russia is one of the clearest examples of an informational autocracy that uses digital technologies to spread state narratives online and involves various digital communities in shaping these narratives. Platforms' participatory dimension allows content aligned with pro-regime messages to emerge outside the state sphere and facilitates the process of authoritarian legitimation. Unlike classic forms of legitimacy arising from democratic consent and legal processes, authoritarian legitimacy is constructed through discourse about power that shapes how citizens reason about the regime and persuades them to accept it (Omelicheva, 2016).

Among different forms of online participatory culture that contribute to authoritarian legitimacy, memes play a special role. The potency of memes is amplified by the engaging nature that makes them usable for communicating (potentially extreme) ideas to the wider public and the intertextuality that makes them understandable without extensive knowledge of the cultural context (Makhortykh & González Aguilar, 2020). During crises, these features are particularly important, as regimes with personalized authority need to mobilize public support continually to sustain legitimacy and to reaffirm the leader's image as a protector (Hutcheson & Petersson, 2016). In this way, memes serve as a mechanism through which digital communities contribute to sustaining authoritarian power beyond official communication.

Within this system of meme-assisted legitimacy co-production, far-right online communities occupy an important role. While research on Russian propaganda and state-controlled media keeps expanding (Geissler et al., 2023; Lankina et al., 2020; Paul & Matthews, 2016), relatively few studies look at how the semi-autonomous far-right actors contribute to sustaining regime legitimacy. In the Russian context, such actors have been selectively co-opted by the state: Laruelle (2015) notes that nationalist and conservative movements have been gradually incorporated into official ideology since the early 2010s, reinforcing Kremlin narratives of patriotism and anti-liberalism. Gaufman similarly shows that Russian far-right serve the Kremlin's agendas, reproducing the discourse of "ethnicity and empire as opposed to class or democracy" (Gaufman, 2025, p. 20).

To understand how far-right communities in Russia may utilize online participatory culture to contribute to authoritarian legitimacy, we examine memes produced by these communities in relation to contemporary politics. By doing so, we aim to address our main research question: How do far-right Telegram channels use Internet memes to construct (de)legitimacy narratives around Russian and foreign political figures?

## Authoritarian legitimacy in the context of Russia

In the early 2000s, Vladimir Putin's legitimacy largely depended on economic recovery and the promise of stability after the turbulent 1990s (Burrett, 2019). As economic growth slowed after the 2011–2012 protests, performance-based legitimacy increasingly gave way to information manipulation (Guriev & Treisman, 2019), with narratives of moral conservatism (Sharafutdinova, 2014), sovereignty, and historical continuity (Malinova, 2022). Miazhevich (2018), with the example of Russia's international multinational broadcaster RT, shows how the Kremlin strategically constructs nationalistic narratives of Russia as a protector of its neighborhoods and Putin as the embodiment of Russia.

Prior research has identified several recurring themes in Kremlin propaganda which serve as a basis for Russia's model of informational autocracy. For example, victimhood and heroism depict Russia as a victim of Western aggression while framing it as a heroic challenger to Western hegemony (Miazhevich, 2018; Tolz & Hutchings, 2023). Another theme regards Russian nationhood (Hutchings & Szostek, 2015) built on ideas of a strong state (Prozorov, 2005), sovereign democracy (Casula, 2013; Evans, 2008), and conservatism (Robinson & Milne, 2017; Sharafutdinova, 2014). The order and stability theme reinforces the regime's image as a guarantor of sovereignty and modernization (Frear & Mazepus, 2021). All these themes often involve portraying the West as morally degraded, hostile, and threatening to Russian sovereignty (Szostek, 2017; Tolz & Hutchings, 2023).

Since the annexation of Crimea, the nationalistic discourse and "rally-around-the-flag" dynamics have become more salient in Russian politics (Kazun, 2016) as the Kremlin uses the growing hostilities with the West to suppress dissent and strengthen public support. These shifts aligned with a discursive change described by Hutchings and Szostek (2015), where communication itself becomes a key means of reinforcing legitimacy (Guriev & Treisman, 2019).

After 2022, the Kremlin faces growing pressure on its legitimacy, as the full-scale invasion of Ukraine undermined the image of competence on which Putin's authority had rested (Stoner, 2023) and increased the need to justify military actions, frame the conflict as a necessity, and reinforce narratives of patriotism, unity, and external threat. In this context, propaganda becomes a central mechanism of domestic legitimation, and its task is harder under a military crisis, which puts particular pressure on an information autocracy. Far-right actors play an important role in post-2022 Kremlin propaganda and, subsequently, the system of authoritarian legitimation, albeit this role is rather ambiguous. While the Russian government is "mainstreaming the far-right discourse" (Gaufman, 2025, p. 28), many communities reiterate Kremlin narratives and push them to the extreme. Some did not support the invasion or later attacked the legitimacy of the regime and faced retaliation, but most remain staunchly loyal.

## Far-right memes as a form of authoritarian legitimacy

The research on memes in political communication has been expanding recently, highlighting their significance for political expression, narrative construction, and activism (Davis et al., 2016; Dimitrov et al., 2021; Makhortykh, 2015; Nielsen & Nititham, 2022; Nørgaard Kristensen & Mortensen, 2021; Peters & Allan, 2022). Shifman defines memes as a group of digital items that have a common characteristic of content, are created in awareness of each other, and are circulated, imitated, and transformed via the internet (Shifman, 2013). Almost every major political event now produces a stream of memes, a pattern Shifman (2014, p. 23) calls "hypermemetic logic" (see also Milner, 2016).

Memes can shape collective identities as they rely on shared cultural references and give a common language for interpreting politics (Milner, 2016). Memes also strengthen ingroup ties and hostility toward outgroups (Milner, 2016). They are therefore a site where the boundary between "us" and "them" is drawn (Mortensen & Neumayer, 2021).

Research on memes in authoritarian contexts has predominantly examined them as means for citizens to criticize authorities, express disagreement, and negotiate identities under conditions of repression (Moreno-Almeida, 2021). However, the same features enabling the use of memes for dissent make them powerful tools for autocratic governments (Makhortykh & González Aguilar, 2020). The Kremlin, for example, appropriates memetic practices to reproduce official narratives and increase popular support (Denisova, 2019; Wiggins, 2016). However, much less attention has been devoted to grassroots pro-autocracy communities.

Far-right communities embed their ideology in casual content spread through internet culture, where humor masks extreme ideas as jokes and makes such content easier to share with mainstream audiences (Fielitz & Marcks, 2019). In this way, memes normalize exclusion and political violence (Fernández-Villanueva & Bayarri-Toscano, 2021). Russian far-right communities adapt global meme formats to Russian content and use them to define their ingroup and enemies (Oskolkov et al., 2026).

The (de)legitimizing role of far-right memes can be understood through the lens of Easton's (1965) concepts of diffused and specific support, which are related to the notion of legitimacy. Specific support relies on perceptions of leaders' competence and authority, making their image management central to political systems' stability, particularly under the condition of weak institutional trust. Ross and Rivers (2017) show that memes about Hillary Clinton and Donald Trump worked as (de)legitimization strategies that targeted specific support by attacking or reinforcing individual competence. The same strategies appear in far-right memes, which on 4chan glorified Trump's authoritarian style and attacked him when he seemed not authoritarian enough (Way, 2021) and in Bolsonaro's campaign legitimized hate and violence against his opponents (Fernández-Villanueva & Bayarri-Toscano, 2021). Because of their provocative and extreme nature, far-right memes circulate rapidly and generate strong emotional engagement, which makes them highly effective for mobilizing authoritarian support and undermining the legitimacy of their opponents.

## Methodology

### Data collection

To implement the study, we collected memes from public Telegram channels associated with pro-Kremlin Russian far-right communities. We chose Telegram because of its widespread use among Russian-speaking audiences (Khaund, 2020), together with minimal content moderation, which facilitates the circulation of extremist and politically charged content. Telegram's role as a tool for both state-aligned actors and semi-autonomous far-right groups makes it a particularly relevant platform for examining how online participatory culture contributes to authoritarian legitimation. We looked at memes published after Russia's full-scale invasion of Ukraine from February 24, 2022, to November 20, 2025, to examine how legitimation strategies function under conditions of crisis.

We focused on memes depicting Vladimir Putin, Dmitry Medvedev, Volodymyr Zelensky, Joe Biden, and Donald Trump. Putin represents the symbolic core of the regime and is expected to be at the core of legitimation efforts. Medvedev illustrates a legitimation mechanism captured by the Russian saying "the Tsar is good, but the nobles are bad", where criticism is redirected toward subordinate elites to protect the leader's authority (Mamonova, 2016). Foreign leaders such as Biden and Zelensky are expected to be targeted by delegitimizing narratives aligned with authoritarian ideological tropes of external threat and Western decay. Finally, Trump represents a strategically ambiguous case due to his contentious role in the world's politics.

To collect memes, we extracted all visual content published between 2022 and 2025 on 31 channels using Telegram's application programming interface (API) (the complete list of channels is provided in Appendix A). An initial set of channels was identified by the authors based on their knowledge of the far-right Telegram meme communities, and additional channels were located through a snowball sampling that traced reposts by initial channels. Altogether, we extracted 404,013 images.

As an initial filtering step, images were processed using the Multi-task Cascaded Convolutional Neural Network (Zhang et al., 2016) implemented via the facenet-pytorch library to detect the presence of human faces. A minimum confidence threshold of 0.6 was applied, and an image was considered to contain a face if at least one detected face exceeded this threshold. Images without detected faces were excluded from further analysis. This step resulted in a reduced corpus consisting of 136,984 images.

### Data analysis

To identify politicians of interest, we used Amazon Rekognition (Amazon Web Services, n.d.). First, we applied the default celebrity recognition API to identify well-known public figures. For Zelensky, who was not covered by the default API, a custom reference collection of images was created. Similarity-based face search was performed against this collection for all images with at least one unrecognized face.

Identical images were identified using MD5 hashing, and one copy of each was kept, including images that appeared several times in the same or in different channels. After removing duplicates, the final dataset comprised 2,783 images featuring Putin, 2,207 featuring Trump, 1,912 featuring Zelensky, 837 featuring Biden, and 513 featuring Medvedev (the numbers before deduplication are given in Table 1).

**Table 1.** Number of images per politician before the removal of identical copies

| Politician | Images |
|---|---|
| Vladimir Putin | 4,772 |
| Donald Trump | 3,618 |
| Volodymyr Zelensky | 2,408 |
| Joe Biden | 1,214 |
| Mikhail Svetov | 981 |
| Dmitry Medvedev | 920 |
| Ryan Gosling | 902 |
| Igor Strelkov | 874 |
| Alexey Navalny | 767 |
| Elon Musk | 690 |
| Emmanuel Macron | 674 |

Note. The counts include all copies of an image posted in the channels. Identical copies, including images that appeared several times in different channels, were removed using MD5 hashing. The numbers after removal are reported in the text.

Following this, image embeddings were generated using a CLIP (Contrastive Language–Image Pretraining) model (Radford et al., 2021). Clustering was performed using the Hierarchical Density-Based Spatial Clustering of Applications with Noise technique to identify groupings of memes based on visual similarity (Allaoui et al., 2020). This resulted in 115 clusters for Putin, 87 for Trump, 88 for Zelensky, 40 for Biden, and 27 for Medvedev. After this step, one of the authors manually reviewed all images within each cluster to identify the most frequent visual theme in each cluster. Model settings and the caption generation procedure are reported in Appendix B, and the translations of the texts on the memes are given below each figure. In some cases, clusters included the same meme template with different captions; other clusters brought together images with a common visual motif, such as politicians in formal suits or military settings. The dominant theme identified through this qualitative inspection served as the basis for assigning each cluster to a broader legitimization or delegitimization narrative associated with a specific politician.

## Results

We started our analysis by examining the 20 most frequent words that were extracted from the automatically generated captions for each politician after preprocessing (Table 2). Putin's corpus was dominated by words related to national identity ("russian"), state authority ("political", "president", "professional"), and diplomacy ("diplomatic", "meeting", "conference"), together with military references ("military", "uniform"). Other common words such as "social", "conversation", and "headline" point to his images often being connected to public communication, while "humorous" reflects a strong ironic component. Similar elements appeared in the corpus for Zelensky, including terms related to identity and military ("ukrainian", "military", "uniform", "jacket"), politics ("political", "government", "conference",

"office", "diplomatic"), public ("press", "headline", "social", "poster") and personal communication ("humorous" and "professional").

**Table 2.** Most frequent nouns and adjectives in image captions per politician

| Rank | Putin | Medvedev | Trump | Biden | Zelensky |
|---|---|---|---|---|---|
| 1 | russian (956) | russian (270) | suit (902) | suit (363) | ukrainian (440) |
| 2 | suit (920) | suit (198) | political (420) | political (152) | military (327) |
| 3 | military (512) | dmitry (99) | russian (350) | russian (146) | suit (287) |
| 4 | political (387) | military (93) | donald (246) | military (113) | russian (215) |
| 5 | humorous (232) | political (59) | american (230) | american (99) | humorous (152) |
| 6 | putin (209) | humorous (37) | trump (220) | humorous (94) | political (143) |
| 7 | social (181) | professional (34) | humorous (206) | joe (75) | jacket (141) |
| 8 | vladimir (176) | medvedev (33) | tie (168) | government (65) | conference (109) |
| 9 | diplomatic (163) | president (33) | office (133) | smile (59) | uniform (93) |
| 10 | professional (135) | conference (32) | diplomatic (128) | biden (58) | professional (92) |
| 11 | arrange (130) | meeting (30) | president (122) | diplomatic (56) | government (85) |
| 12 | headline (127) | nuclear (29) | map (121) | president (54) | ukraine (82) |
| 13 | president (126) | social (28) | united (117) | professional (47) | social (63) |
| 14 | meeting (126) | conversation (27) | meeting (110) | united (45) | sky (62) |
| 15 | uniform (122) | symbol (27) | military (109) | jacket (43) | office (62) |
| 16 | portrait (116) | russia (27) | social (104) | ukrainian (42) | contemplative (61) |
| 17 | engage (112) | government (27) | smile (100) | social (41) | diplomatic (60) |
| 18 | russia (111) | office (25) | oval (96) | engage (38) | headline (60) |
| 19 | conference (110) | follow (24) | composite (90) | gesture (37) | poster (58) |
| 20 | conversation (107) | poster (24) | ukrainian (90) | office (37) | press (57) |

Note. Word frequencies are in parentheses. Captions were generated with Qwen2-VL-2B-Instruct and lemmatized. Only nouns and adjectives were kept, and stopwords were removed. Images not assigned to a cluster were excluded.

Biden's corpus reflects his role as U.S. president with words related to state and diplomacy ("government", "diplomatic", "president", "office", "united"). Strikingly, "Russian" outranks "American" in his top words, suggesting that Biden is captioned predominantly in relation to Russia. "Smile" also ranks highly, corresponding to how Biden is shown in most of the images. Trump's corpus similarly was dominated by politics-related terms, but with a more formal emphasis: he is more often placed "in office" ("office", "oval", "diplomatic", "president"). As with Biden, "Russian" and "American" sit close together in ranking, positioning Russia as a recurring element of Trump's representation. An ironic component is also present ("humorous", "smile"), and "composite" suggests the presence of modified images. Medvedev's corpus is centered on identity and politics ("political", "government", "conference", "meeting", "office"), also referencing military, including nuclear weapons (“military”, “nuclear”). A communication dimension is also prominent for Medvedev (“social”, “conversation”, and “follow”).

The obtained results describe general thematic patterns, but they are insufficient for identifying nuanced aspects of representation, such as irony or visual metaphors, because the generated captions describe only the literal aspects of memes. It also cannot distinguish between different representations emerging from similarly described captions: Putin’s and Zelensky’s corpora share similar keywords, but their visual composition is very different. This stresses the need to combine a quantitative approach with more qualitative reading of memes that we provide below.

**Partial delegitimization: Medvedev is an alcoholic and a radical version of the President**

Dmitry Medvedev occupies a semi-legitimate position within far-right meme discourse. He remains symbolically connected to the Russian political elite and continues to be represented as a politically recognizable figure, yet he is rarely portrayed as an actor having real political agency. The memetic reiteration of his legitimacy derives primarily from his institutional proximity to power rather than competence or decision-making capacity. Such proximity is often shown via official photographs (e.g., of business meetings) and images from Medvedev’s presidential period. He is usually shown wearing a suit or a military outfit, often together with Putin (e.g., discussing geopolitical decisions) but rarely taking any action. This closeness to Putin (Figure 1) legitimizes Medvedev, because Putin represents the ultimate form of political agency and leadership in Russian far-right memes.

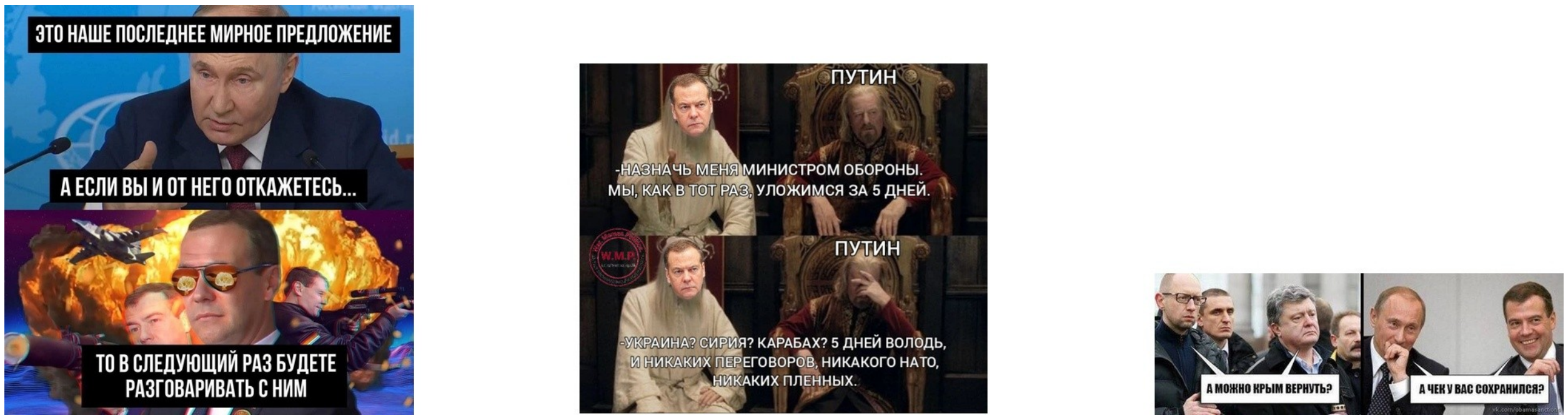

**Figure 1.** Memes focused on Medvedev’s closeness to Putin.

*Translation of the texts on the memes:* “This is our last peaceful proposal… next time you will be talking to him”; “Appoint me minister of defense, we will be done in five days”; “Can we have Crimea back?” “Have you kept the receipt?”

Besides closeness to Putin, Medvedev’s image is also shaped by his provocative behavior online. Multiple memes focus on his social media activity, framed as exaggerated performative escalation. Medvedev’s statements are presented as memes in themselves (Figure 2), transforming his statements into meme material and shifting his political role from governance to the production of offensive entertaining content.

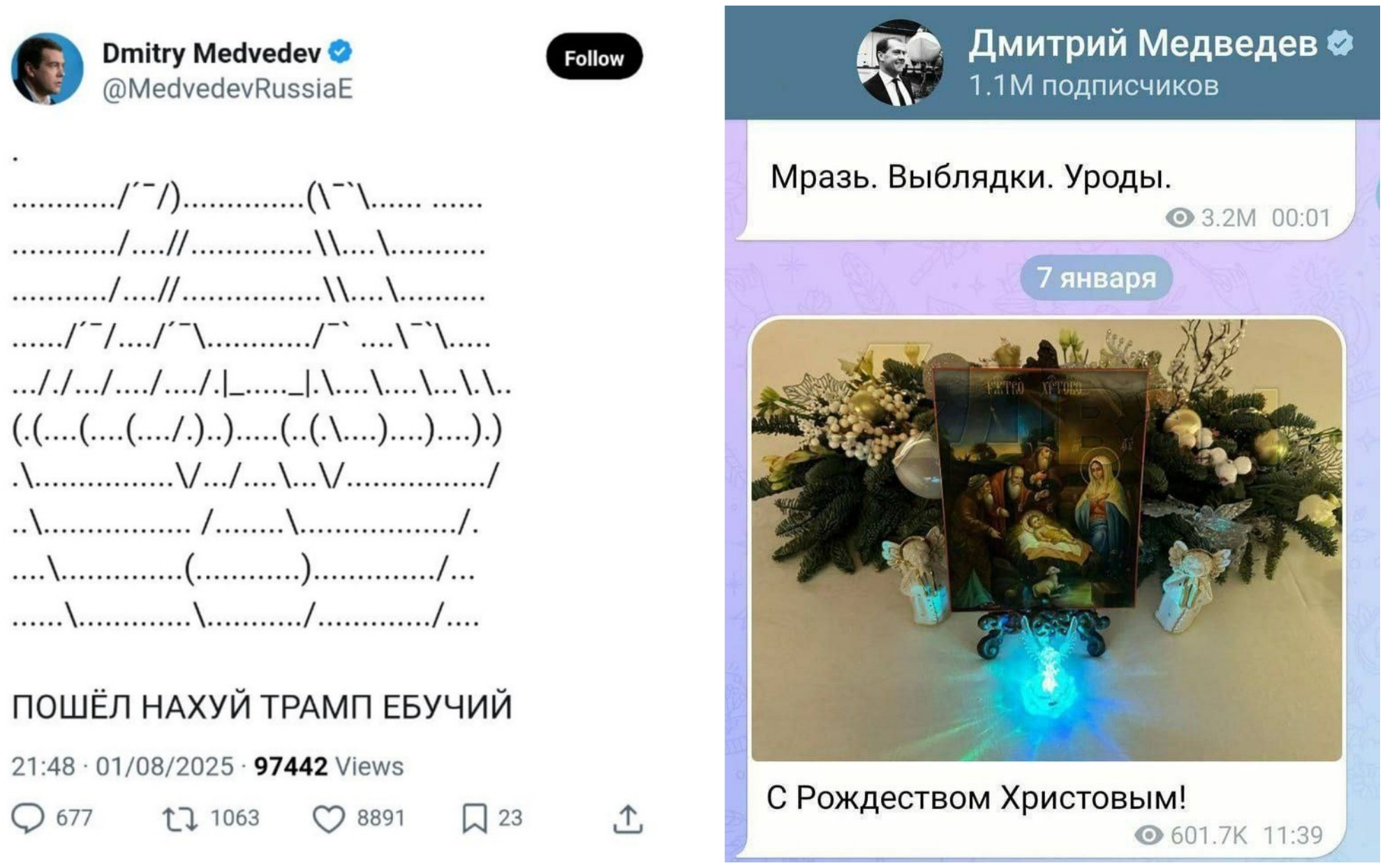

**Figure 2.** Memes focused on Medvedev's social media statements.

*Translation of the texts on the memes:* "Fuck you, fucking Trump"; "Scum. Bastards. Freaks." "Merry Christmas!"

A related group of memes relates to Medvedev's militarized and hyper-aggressive rhetoric connected to the war in Ukraine (Figure 3). His rhetoric does not seem to have a real influence over the course of the war or the decisions of the Russian government, but it carries a symbolic degree of legitimacy, as Medvedev remains associated with the Russian political elite. Medvedev functions as a figure through whom extremist ideas can be articulated, which contributes to the dehumanization of Ukraine and the West. And humor makes this extremist rhetoric easier to circulate and normalize (Askanius, 2021; Schmid & Greipl, 2025).

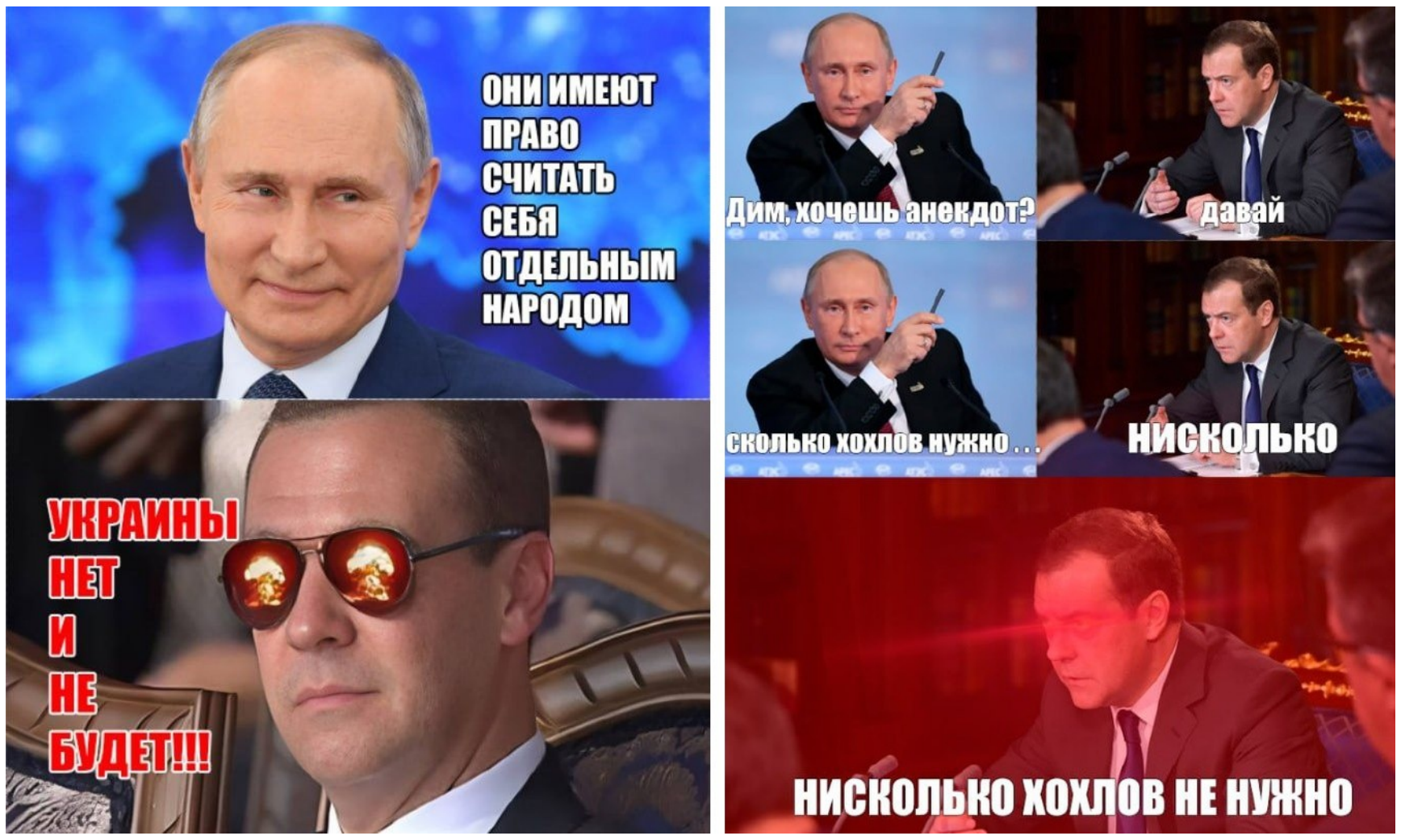

**Figure 3.** Memes focused on Medvedev's militarized rhetoric.

*Translation of the texts on the memes:* "They have the right to consider themselves a separate nation." "There is no Ukraine and there never will be!"; "How many khokhly [a slur for Ukrainians] does it take…" "None."

At the same time, some memes focus specifically on the delegitimization of Medvedev by stripping him of his full political agency, for instance, by referring to his alleged alcoholism (Figure 4). Often, his online behavior is presented as the result of intoxication, which shows Medvedev as rather incompetent and weak. This representation undermines his status as an influential political actor and indirectly reinforces the legitimacy of Putin, who presumably saw Medvedev as one of his competitors in the 2000s.

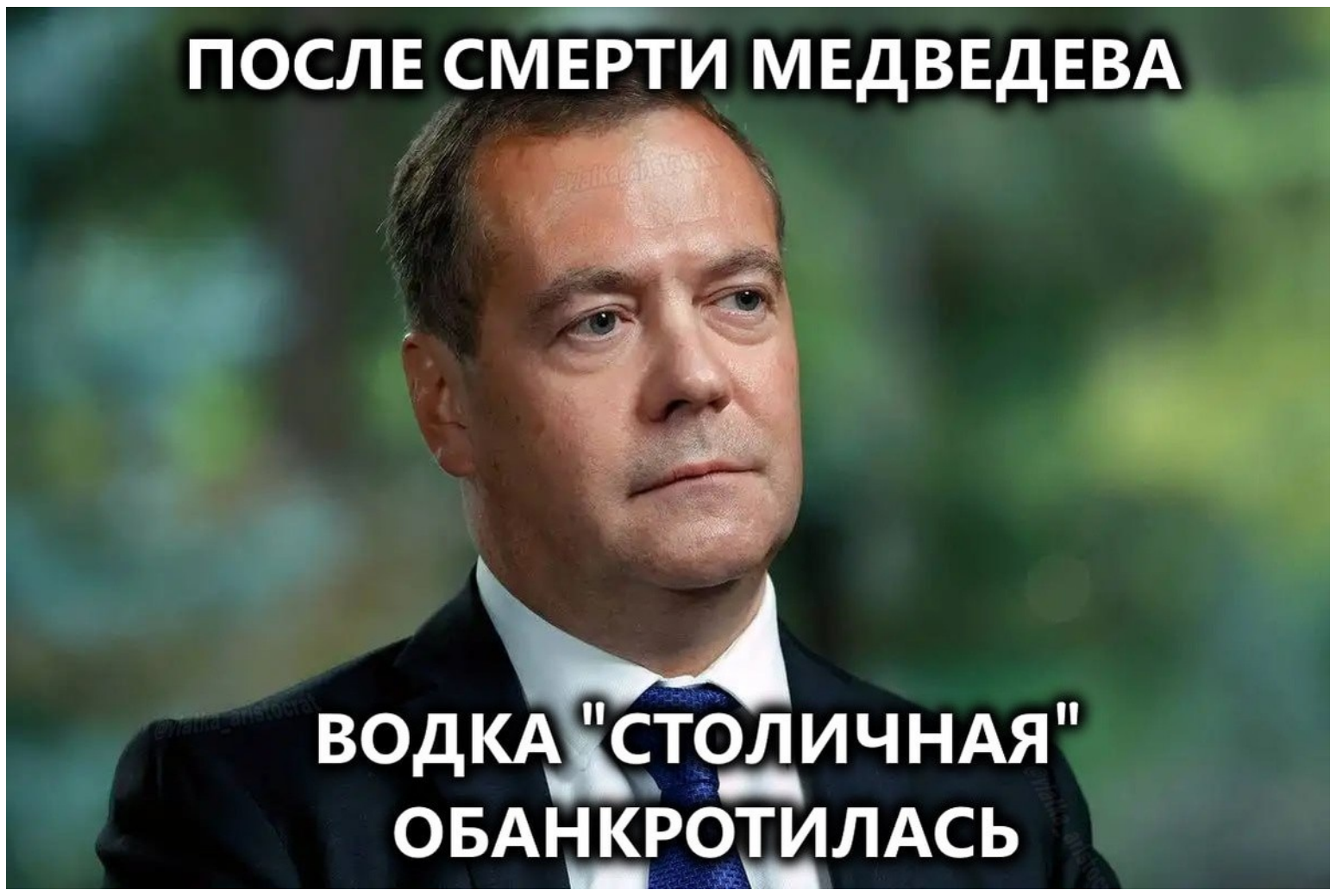

**Figure 4.** Meme focused on Medvedev’s alleged alcoholism.

*Translation of the texts on the memes:* “After Medvedev’s death, Stolichnaya vodka went bankrupt.”

The representation of Medvedev reflects the broader political logic where the radical aspects and failures of the regime are displaced onto subordinate elite figures while the supreme leader remains associated with sovereign authority. Medvedev becomes a subordinate figure onto whom radical escalation can be displaced, protecting Putin from direct association with the most extreme discourse and yet giving a space to share those radical ideas.

### Partial delegitimization: Biden as a Ukrainian ATM with dementia

The representation of Biden is characterized by a consistently delegitimizing discourse. As a president, he is rarely associated with charisma, strong leadership, or geopolitical power; instead, his representation concerns old age, cognitive decline, and political ineffectiveness. At the same time, unlike completely delegitimized Zelensky, Biden is treated as an actor whose decisions still carry some weight in global politics.

The dominant pattern across Biden-related memes is the portrayal of him through the lens of old age (Figure 5), focusing on his mental and physical weakness, including alleged inability to speak coherently. Memes present him as incapable of leadership and systematically undermine his competence. Age becomes the central interpretative framework, transforming Biden into a symbol of Western weakness.

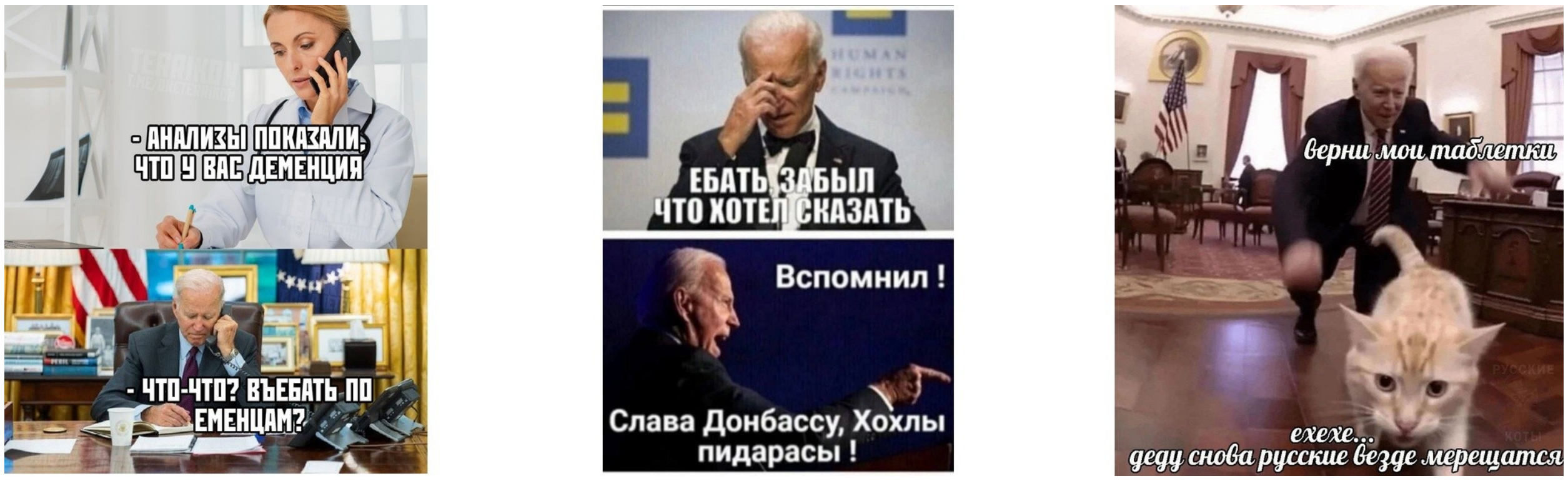

**Figure 5.** Memes focused on Biden's age and cognitive decline.

*Translation of the texts on the memes:* "The tests showed you have dementia." "Bomb the Yemenis?"; "I forgot what I wanted to say. Remembered! Khokhly are faggots!"; "Grandpa is seeing Russians everywhere again."

A recurring pattern is a comparison between Biden and Trump (Figure 6). Memes often contrapose them in a single image, while delegitimizing memetic practices aimed at Biden strengthen Trump's image as a more decisive and authoritative actor. Nevertheless, within these comparisons, Biden appears as a professional actor, wearing a suit and interacting with other political leaders on the same geopolitical level. Although his representation is highly ironic, he still retains a certain degree of legitimacy, as his decisions are shown to influence the course of the war.

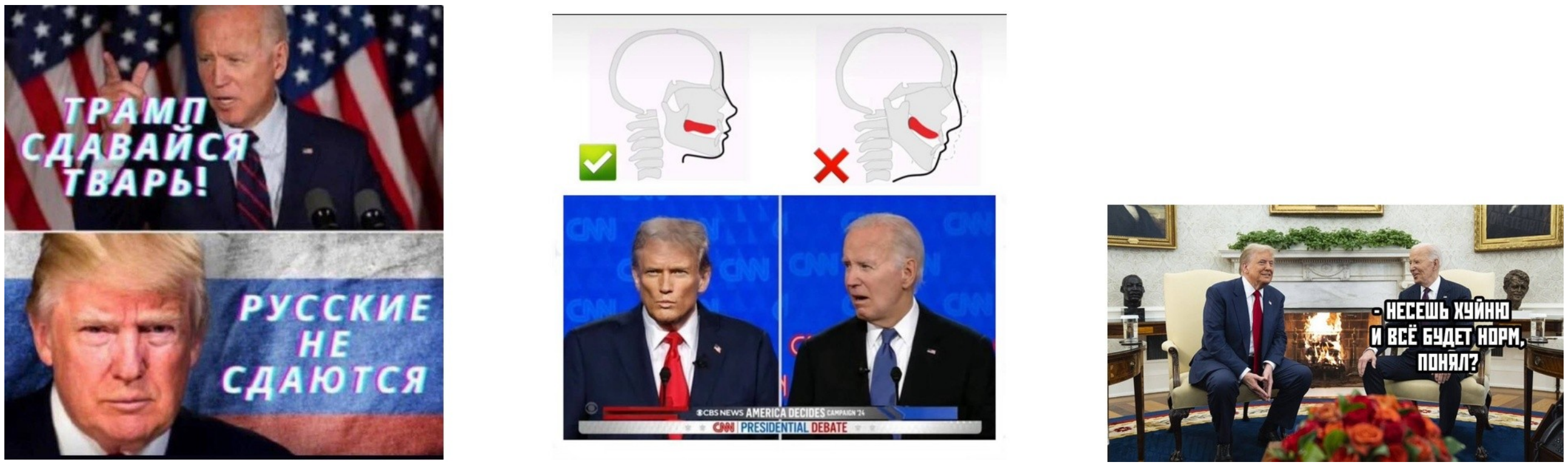

**Figure 6.** Memes comparing Biden and Trump.

*Translation of the texts on the memes:* "Trump, surrender, you scum!" "Russians don't surrender"; "You talk bullshit and everything will be fine, got it?"

A big share of images focuses on picturing Biden with Zelensky (Figure 7), who is treated as the ultimate symbol of oppression and illegitimacy, resulting in Zelensky's "aura of illegitimacy" being transferred to Biden as well. This interaction often concerns Ukraine's requests for financial or military support. Memes construct Biden as a military sponsor of Ukraine — a figure who controls resources and has the authority to satisfy or deny Zelensky's demands. Some memes introduce homoerotic themes into their relationship, framing the political alliance between the two leaders as something shameful from the perspective of traditional values, with Biden portrayed as the dominant figure.

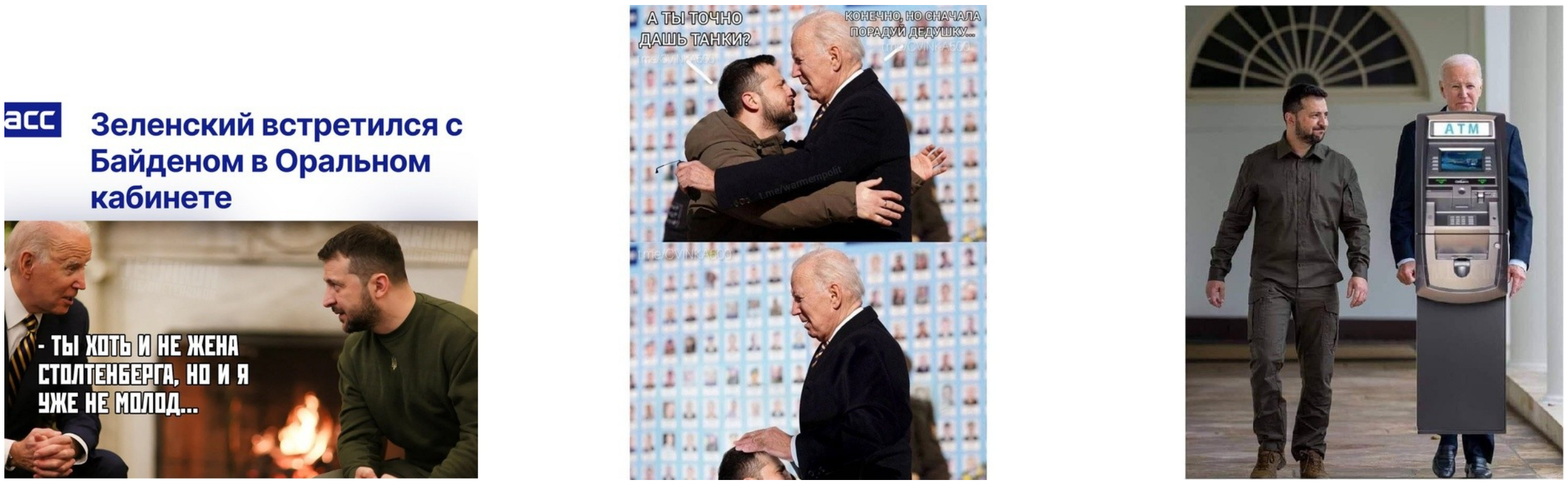

**Figure 7.** Memes focused on Biden's relationship with Zelensky.

*Translation of the texts on the memes:* "Zelensky met with Biden in the Oral Office"; "Will you really give me the money?" "First make grandpa happy".

Another delegitimizing pattern links Biden to drug addiction (Figure 8), where he is connected to drug abuse associated with his son and Zelensky. Memes often accuse Zelensky of cocaine addiction and, by association, extend it to Biden to undermine his ability to rule. In a sense, this strategy is similar to the one employed to delegitimize Medvedev, except for the different source of addiction.

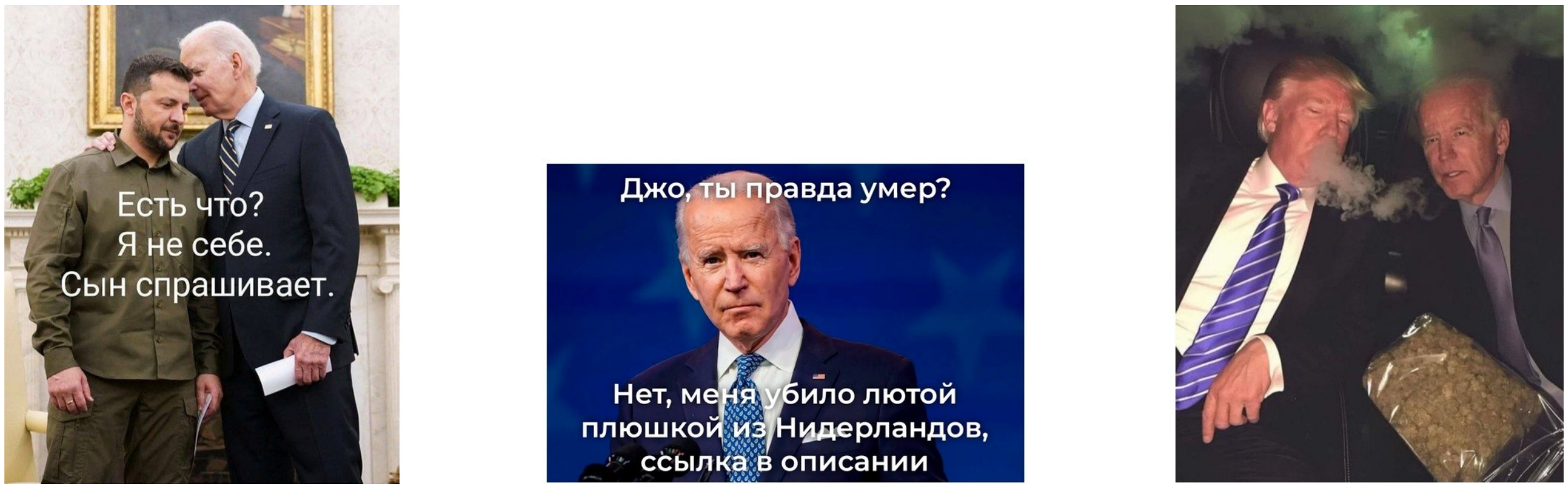

**Figure 8.** Memes linking Biden to drug addiction.

*Translation of the texts on the memes:* "Got anything? It's not for me. My son is asking"; "Joe, did you really die?" "No, I was killed by wicked hash from the Netherlands."

Overall, Biden relatively rarely appears in memes on his own; more often, he is featured alongside other political figures, through whom his image is legitimized or delegitimized. This "secondary character" status highlights his lack of agency, but he is still portrayed as an actor capable of influencing international politics even when this very portrayal is subjected to harsh criticism.

### Selective legitimization: Trump is Putin's American boyfriend and a peacemaker

Trump is another American president central to far-right memes. He is not delegitimized as a representation of the collective West, instead, memes portray him positively, as negotiable and, in many cases, even symbolically close to Russia.

Many memes depict Trump together with Zelensky in peace negotiation settings, presenting Trump as trying to make it on Russian terms (Figure 9). Unlike Biden, Trump is portrayed as someone who resists Zelensky's demands and humiliates him, this contributes to his legitimation in the same way as happens in the case of Russian politicians. Trump's authority is bolstered by his efforts to undermine Zelensky's legitimacy and to appear as an independent politician capable of breaking with the established Western policies, including the unconditional support for Ukraine.

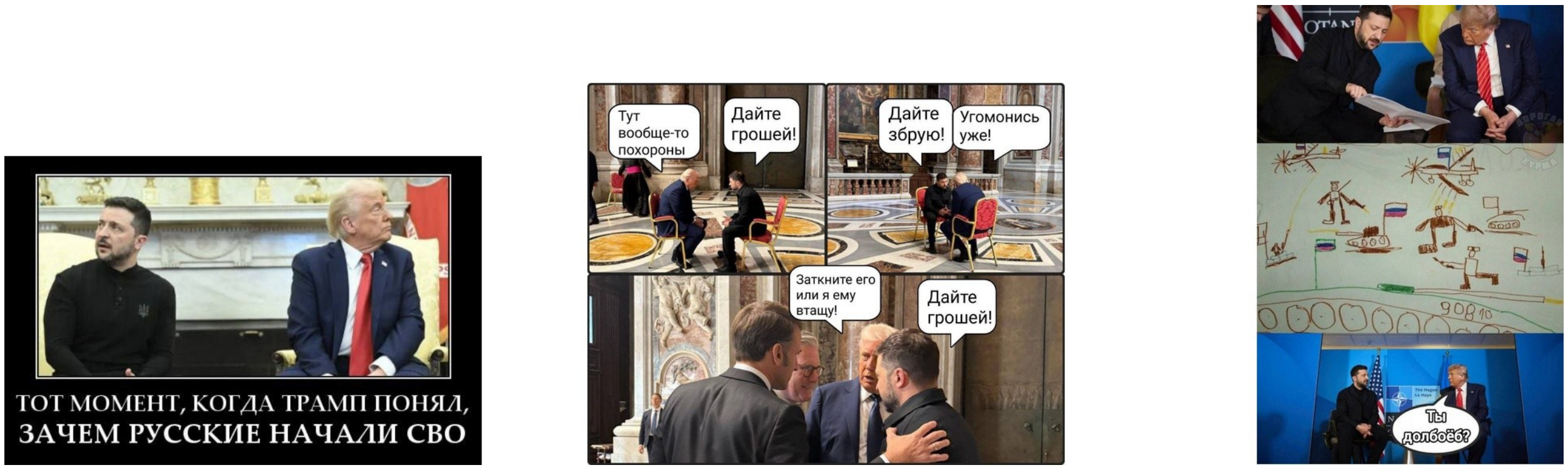

**Figure 9.** Memes depicting Trump and Zelensky in peace negotiations.

*Translation of the texts on the memes:* "That moment when Trump realized why the Russians started the war"; "Give me money!" "Give me weapons!" "Shut him up or I'll punch him!"; "Are you an idiot?"

Trump is also frequently depicted alongside Putin in negotiation settings, but in a completely different role (Figure 10). Their interaction is portrayed as a dialogue between equal actors capable of understanding one another. In these memes, Trump and Putin discuss the peace deal without Zelensky, reinforcing the idea that meaningful geopolitical decisions are made between "real" leaders, presenting Trump as one of the few legitimate Western actors precisely because of his understanding of the Russian idea of war.

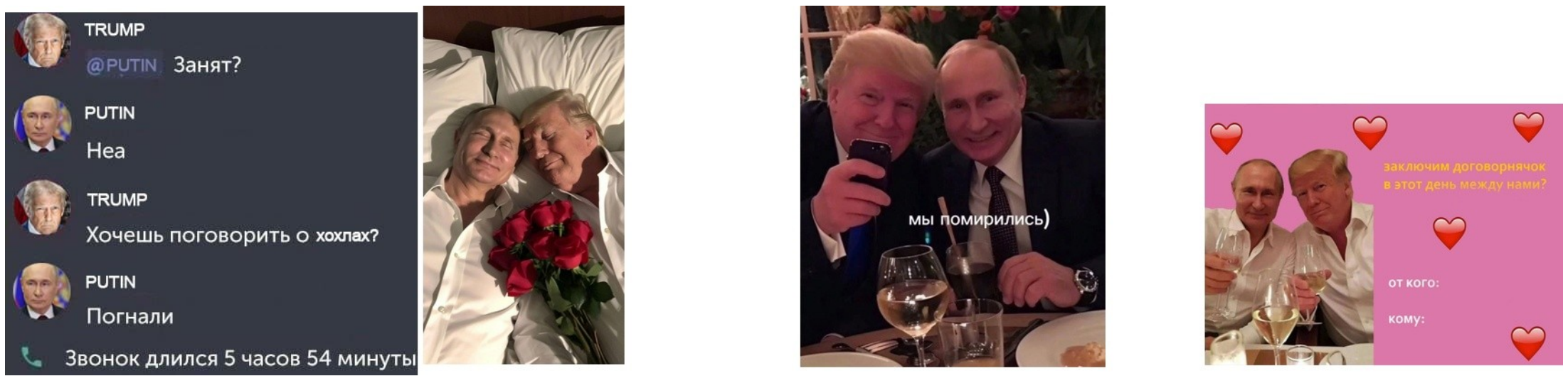

**Figure 10.** Memes romanticizing the relationship between Trump and Putin.

*Translation of the texts on the memes:* "Want to talk about the khokhly?" "The call lasted 5 hours 54 minutes." "We made up :)"; "Shall we make a dogovornyachok [a little deal] between us on this day?"

This logic is supported by memes romanticizing Trump's relationship with Putin and depicting them as partners (or even lovers) who go on dates or exchange valentines (Figure 10). On one hand, such representations contradict traditional masculinity propagated by far-right groups. On the other, they underscore the profound mutual understanding between Trump and Russia, which is expressed metaphorically through his ironic intimacy with Russia's main symbol of legitimacy — Putin (Figure 11).

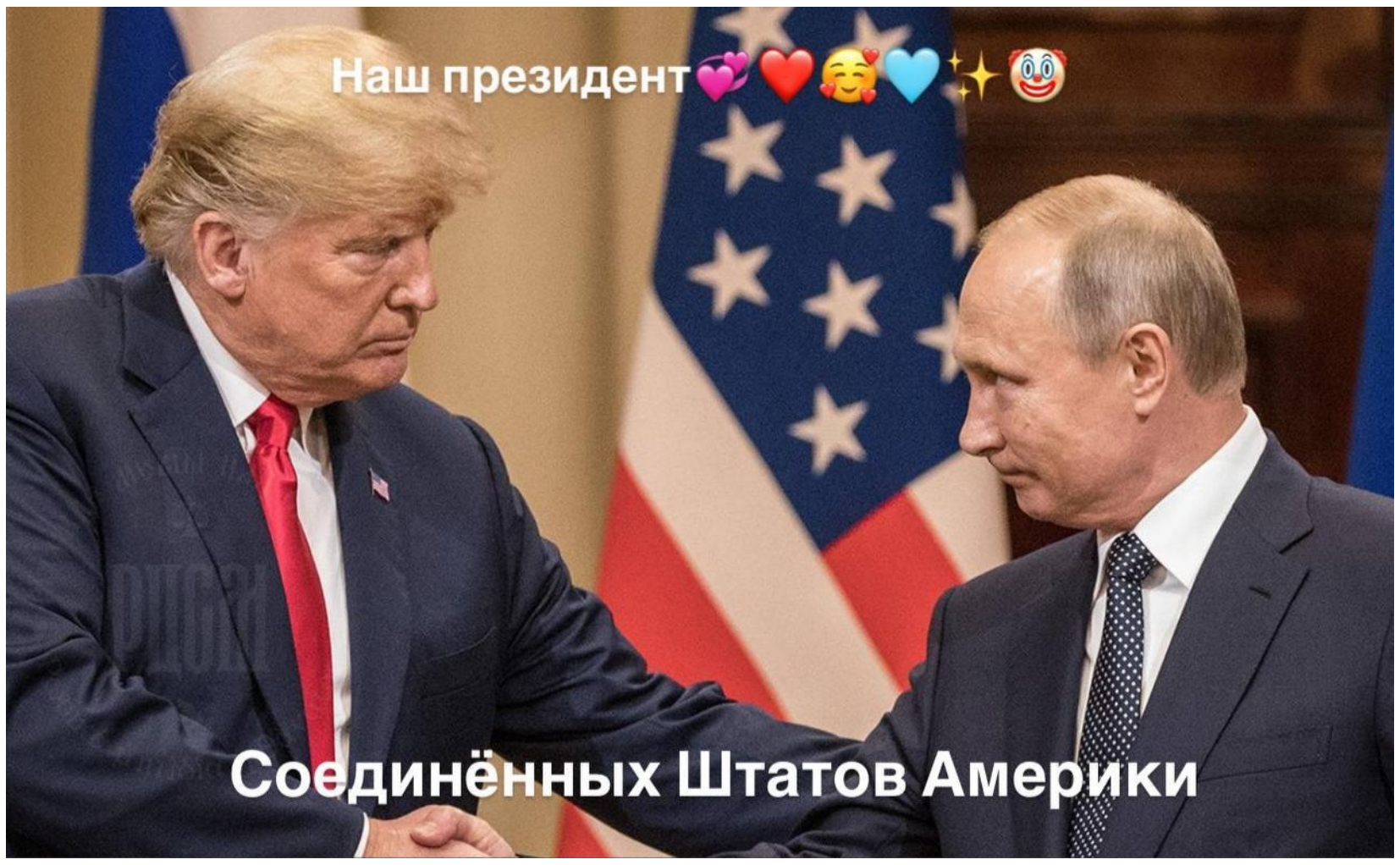

**Figure 11.** Meme depicting Trump as an ally of Putin.

*Translation of the texts on the memes:* "Our president of the United States of America."

Interestingly, Trump is also the only non-Russian political figure who appears in Warhammer-themed memes (see more in the section on Putin), while figures like Biden and Zelensky are not represented there, even as oppressed. The use of Warhammer imagery reinforces the perception of Trump as a strong, authoritarian geopolitical player capable of engaging with Russia from a position of shared views and values.

This closeness between Trump and Russia is further reinforced by the localization of Trump's image in Russian cultural settings (Figure 12). He is depicted as a stereotypical Russian grandmother wearing a headscarf or placed in Russian everyday environments like "Vkusno i tochka" restaurants. Some election-related memes portray Trump as a candidate within the Russian e-government system "Gosuslugi," implying that Russian users themselves would vote for him during the elections.

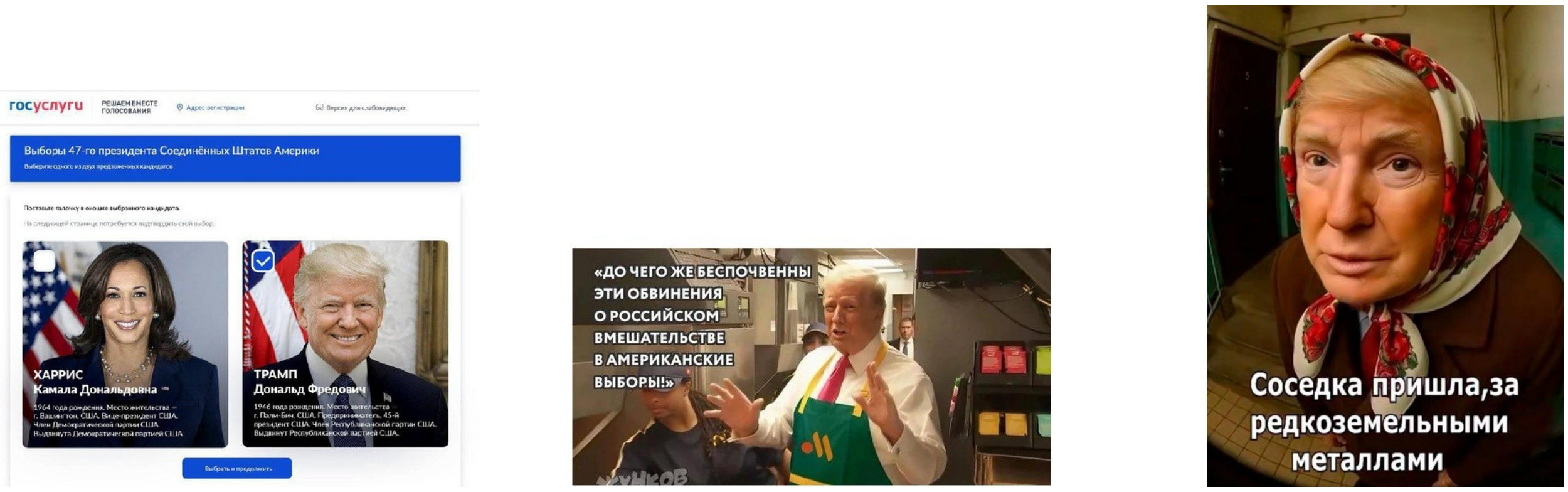

**Figure 12.** Memes localizing Trump in Russian cultural settings.

*Translation of the texts on the memes:* "Election of the 47th President of the United States"; "How groundless these accusations of Russian interference are!"; "The neighbor came over for rare-earth metals."

A related cluster of memes depicts Trump with other actors considered friendly to Russia, such as Orban or Xi Jinping, and portrays him as one of Russia’s historical leaders (e.g., Stalin; Figure 13). At the same time, memes emphasize Trump’s American identity with MAGA symbols and “America is back” slogans, highlighting nationalism and anti-liberal rhetoric. Instead of criticizing Trump as a representative of the "collective West," the creators of these memes seek to find common ground and a sense of symbolic affinity with Russian identity.

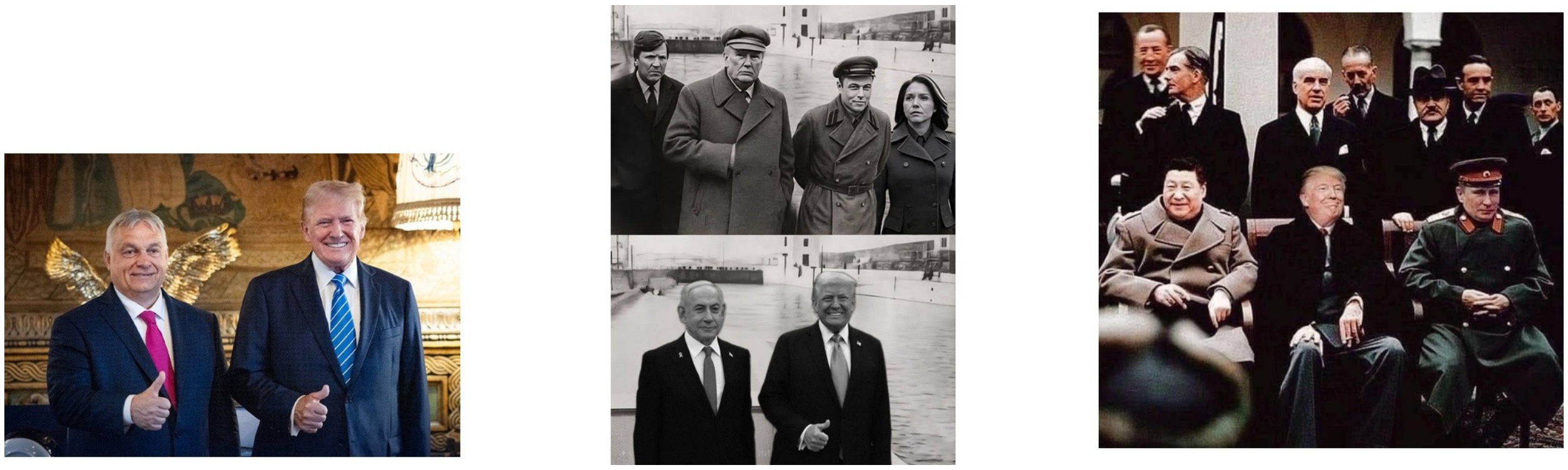

**Figure 13.** Memes depicting Trump with leaders friendly to Russia and with historical leaders of Russia.

More critical memes focus on Trump’s involvement in geopolitical conflicts (Figure 14). While he is shown as an actor attempting to negotiate even in conflicts where he is not a direct participant, many memes mock him as a warmonger. But it also positions him closer to Russia, as the United States is depicted as being engaged in forms of military action typically attributed to Russia itself. Other forms of criticism are focused on Trump’s domestic issues, such as the Epstein affair or failures with tariff introduction. Together, these memes portray Trump as a political figure who makes loud promises without real outcomes; however, this criticism remains mild and is rather for the purpose of irony and not delegitimization.

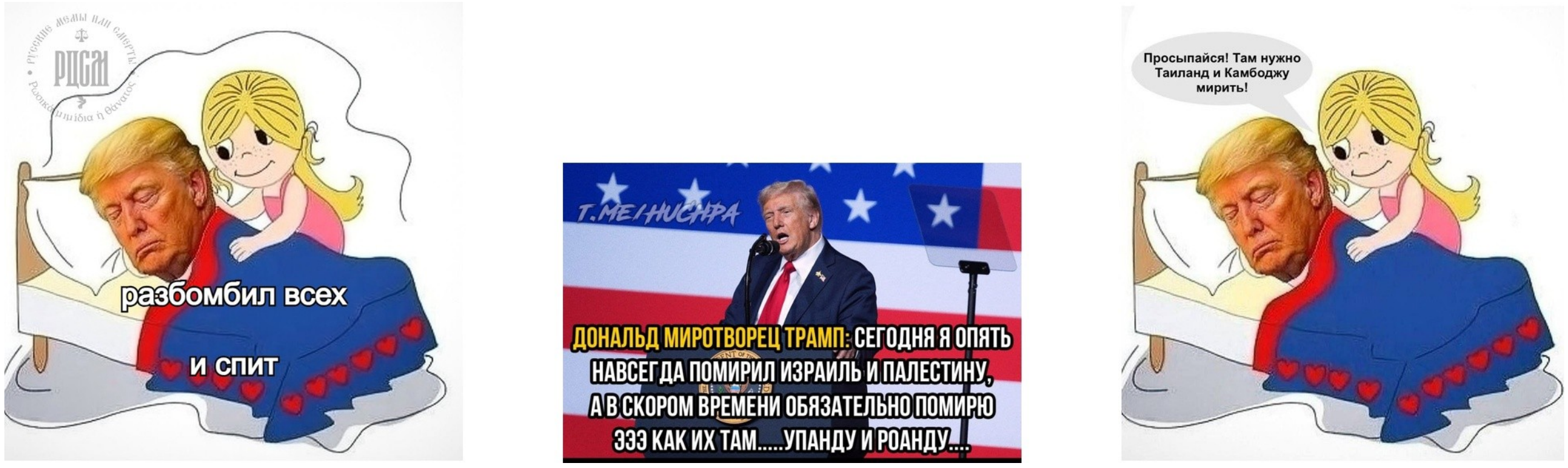

**Figure 14.** Memes criticizing Trump’s involvement in geopolitical conflicts.

*Translation of the texts on the memes:* “Bombed everyone, and now he’s asleep”; “Today I once again reconciled Israel and Palestine, and soon I will reconcile… Upanda and Roanda”; “Wake up! Thailand and Cambodia need reconciling!”

Trump’s representation across the identified topics constructs a positive image of an authoritative actor with probably the most diverse composition of themes observed. While irony and criticism are present, they coexist with broader narratives portraying Trump as capable of negotiation, compatible with Russian geopolitical logic, and in many cases closer to Russia than any other Western political leader.

### Complete delegitimization: Zelensky is a Jewish Nazi clown

The visual representation of Volodymyr Zelensky is characterized by consistently delegitimizing discourse that excludes any even ambiguously favorable portrayals. It relies on strategies that target both his political status and personal qualities. One of **the** key strategies relies on presenting him as an illegitimate leader of a "Nazi regime" responsible for sacrificing his own people in the war (i.e., "killing Nazis" himself; Figure 15). Extreme forms of delegitimization involve direct comparisons with Hitler and Mussolini, which are **complemented** by other offensive, anti-Semitic memes focusing on Zelensky's Jewish identity.

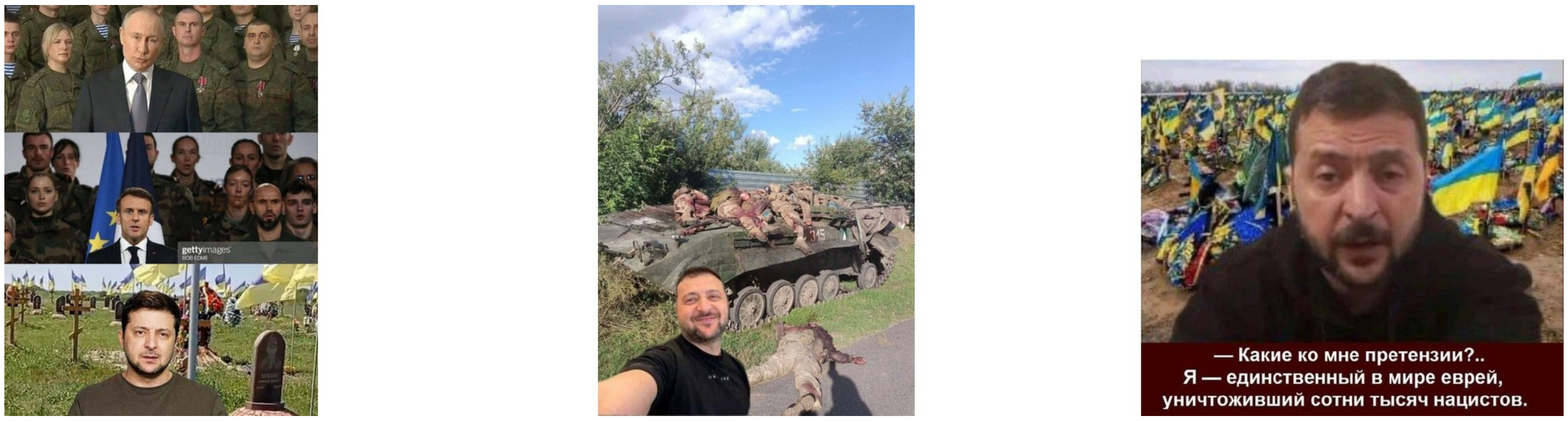

**Figure 15.** Memes presenting Zelensky as the leader of a "Nazi regime."

*Translation of the texts on the memes:* "I am the only Jew in the world who has destroyed hundreds of thousands of Nazis."

Many memes present Zelensky as a subordinate figure who constantly requests financial and military aid from the West (Figure 9). This framing transforms Zelensky from a sovereign political actor into a dependent client of more powerful states who is incapable of sustaining his country's needs. Similar dynamics appear in peacemaking memes on "dogovornyachok," a kind of behind-the-scenes agreement between Trump and Putin. Both politicians emphasize their readiness to establish peace, while Zelensky is portrayed as the sole figure standing in their way (Figure 16). Such representation shifts the responsibility for the continuation of the conflict and human casualties onto Zelensky personally while also undermining his political agency.

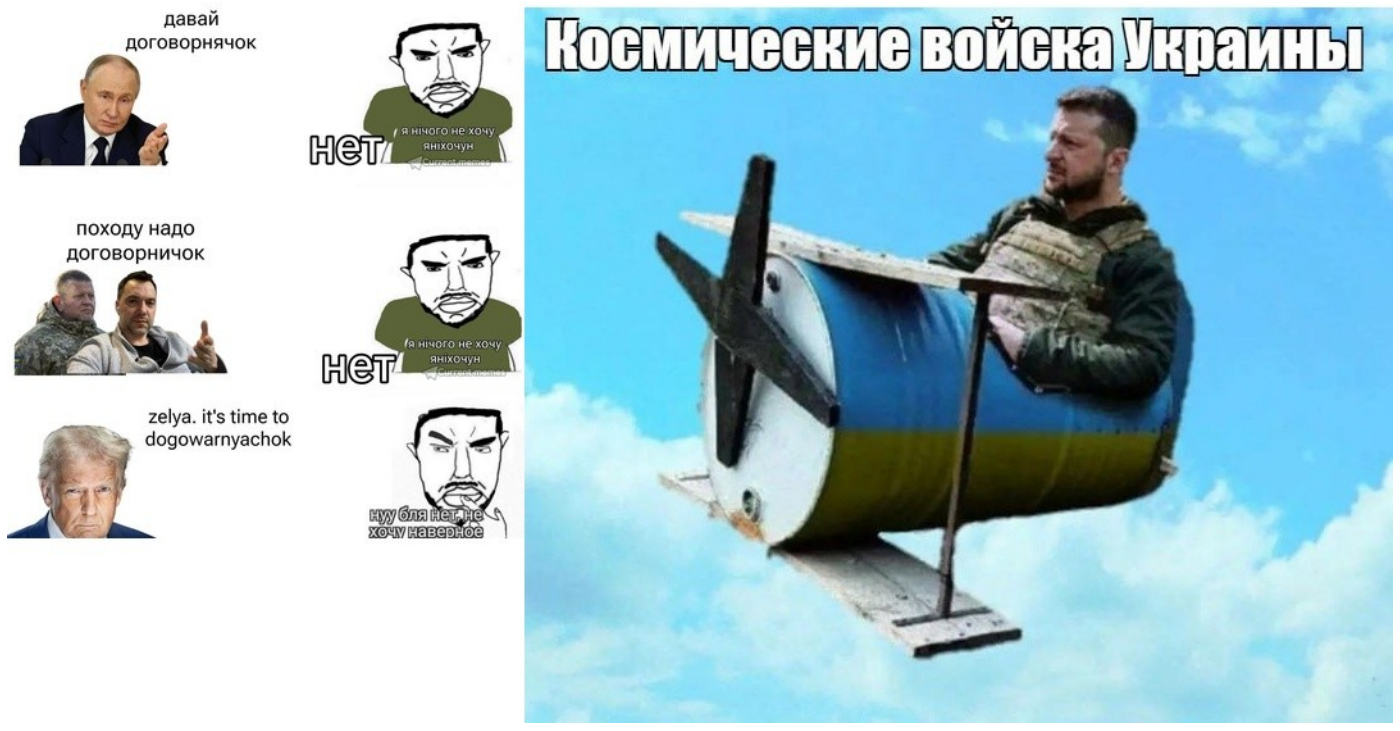

**Figure 16.** Memes focused on Zelensky's lack of agency and the Ukrainian military.

*Translation of the texts on the memes:* "Let's do a dogovornyachok [a little deal]." "No."; "Zelya, it's time to dogovornyachok." "I probably don't want to."; "N-no… you can't…"; "Space Forces of Ukraine."

This absence of agency (or its occasional abuse) is further reinforced through imagery of Ukrainian military incompetence. Memes frequently depict the Ukrainian army as absurdly dysfunctional, for example, showing Zelensky in improvised aircraft made from rusty water barrels (Figure 16). These memes show that a leader who cannot organize the defense of the state fails at the core task of leadership, which undermines the perceived competence on which his legitimacy rests.

Another major dimension of delegitimization involves the degradation of Zelensky's personal image (Figure 17). Some memes portray him as a clown, but the most recurrent trope regarding Zelensky is his supposed drug addiction. Memes portray him as irrational and unstable, implying that his behavior is driven by substance abuse and therefore cannot be trusted to make decisions regarding war and peace; In this way, he is stripped of the rationality expected of a legitimate authority figure. In the same vein, other memes portray him as mentally ill, for example, claiming that he communicates with aliens.

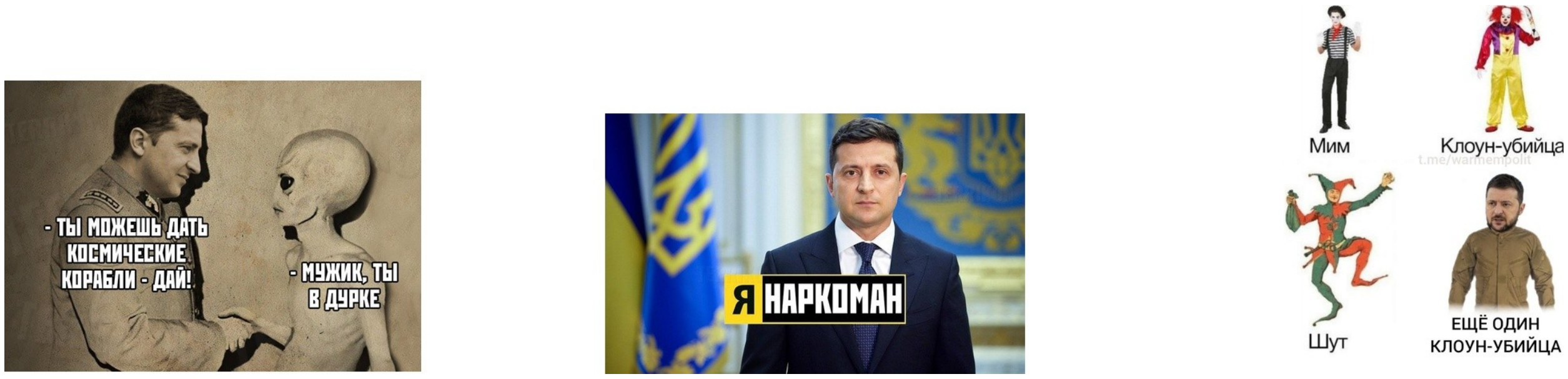

**Figure 17.** Memes focused on Zelensky's personal image.

*Translation of the texts on the memes:* "Give us spaceships!" "Man, you're in a psych ward"; "I'm a drug addict"; "Mime. Killer clown. Jester. Another killer clown."

As part of personal delegitimization, Zelensky is frequently depicted as feminine or engaged in an intimate relationship with Western leaders. A recurring example is the image of Zelensky on high heels and in latex clothing as part of the parody performance (Figure 18). Within far-right Telegram discourse, such imagery is associated with degeneracy that contrasts with ideas of masculinity viewed as core attributes of a political leader. Feminized representations of Zelensky, for instance, in wigs and together with a drag persona of Andriy Danilko (i.e., Verka Serduchka), highlight his complicity with “Western decay” and inability to proudly represent or defend his nation.

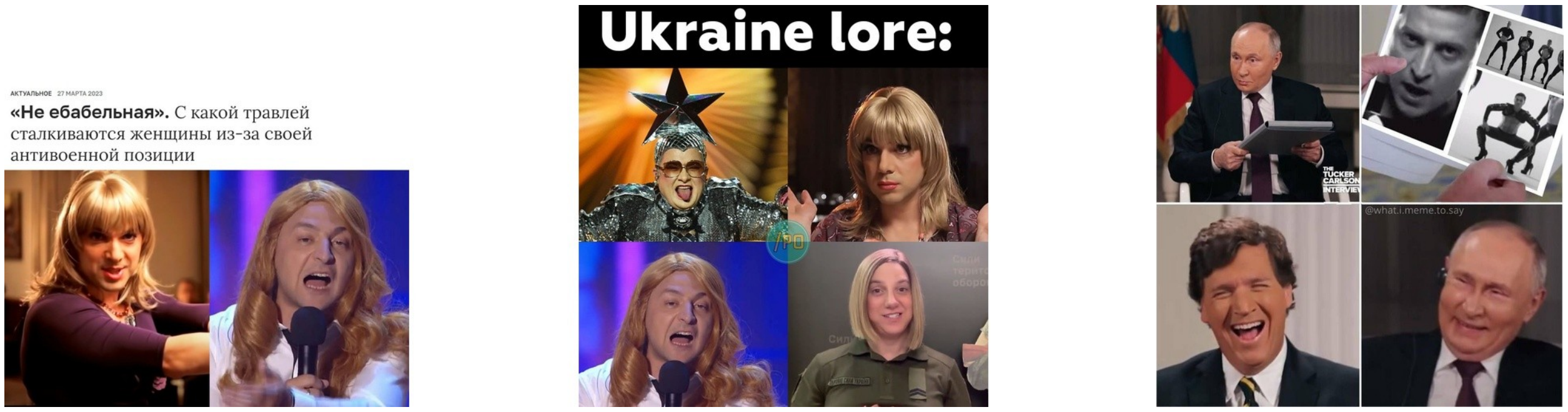

**Figure 18.** Memes depicting Zelensky as feminine.

*Translation of the texts on the memes:* “‘Unfuckable’: the harassment women face for their anti-war stance.”

The most extreme forms of delegitimization rely on direct dehumanization of Zelensky and Ukrainians (Figure 19). Some memes insert Zelensky into films and cartoons, most notably *Heart of a Dog*, where he is presented as Sharikov, a character of a dog that was transformed into a human but remained animal-like inside. In line with this logic, Zelensky is referred to as “Biden’s dog,” emphasizing subordination and lack of autonomy. This dehumanizing representation extends to animalistic imagery applied to Ukrainians, shown as pigs, to strip them of fundamental human qualities.

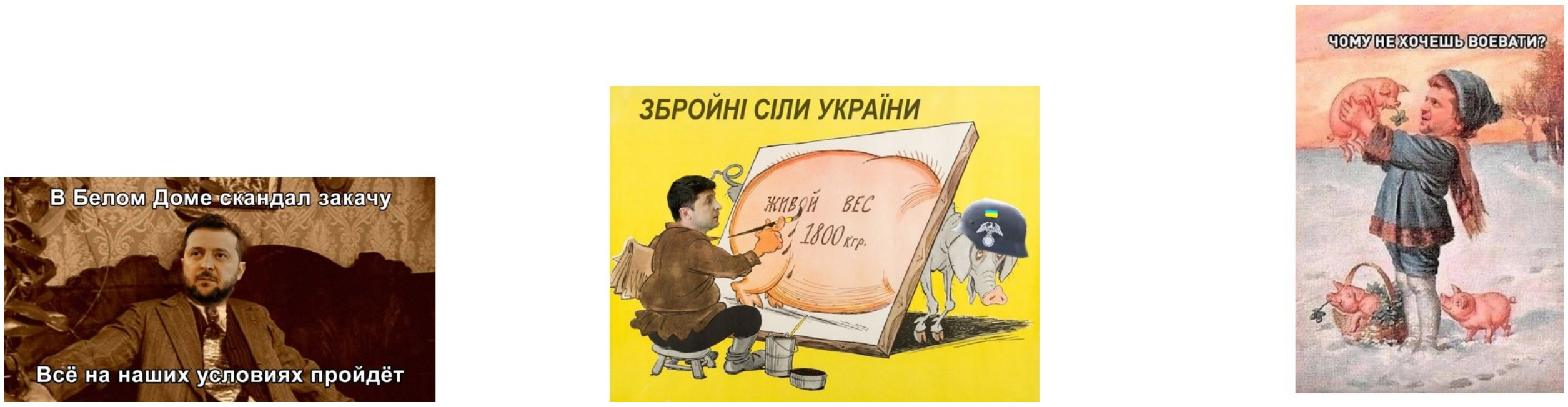

**Figure 19.** Memes dehumanizing Zelensky and Ukrainians.

*Translation of the texts on the memes:* “I’ll make a scene at the White House”; “Armed Forces of Ukraine. Live weight 2,800 kg”; “Why don’t you want to fight?”

Memes about Zelensky are aimed at completely dismantling his ability to lead a country. Humor operates as a mechanism of delegitimization (and often dehumanization), which deprives Zelensky of the qualities associated with legitimate leadership, including sovereignty, rationality, and even humanity.

### (Almost) complete legitimization: Putin is a God-Emperor (but sometimes a monkey)

The visual representation of Putin is built on a legitimizing discourse. He is the central political figure in the memes, appearing more frequently than any other actor. While elements of irony are present, they rarely undermine his status and political power.

A central pattern involves portraying Putin as a legitimate president of Russia. Putin appears in a formal suit, creating a visible contrast with Zelensky, whose military-style clothing is repeatedly mocked. Some groups of memes reproduce pro-Kremlin media aesthetics (Figure 21), aligning with the expected logic of informational autocracy with a competent leader.

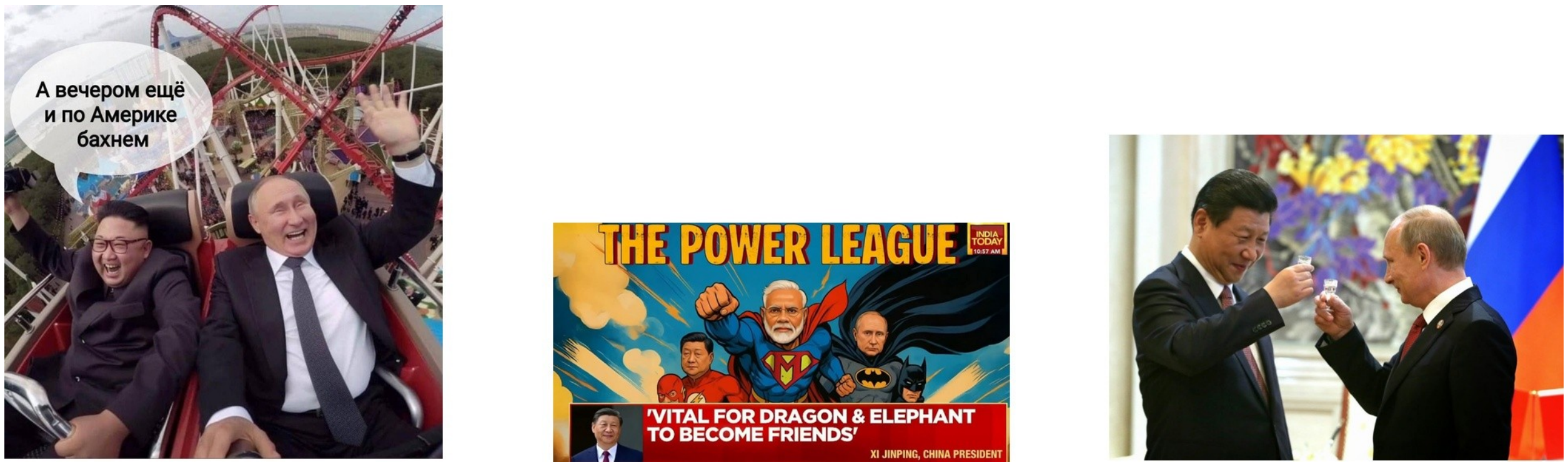

**Figure 20.** Memes depicting Putin with foreign leaders.

*Translation of the texts on the memes:* “And in the evening, we’ll hit America.”

Memes also shape the image of Putin as a head of state who possesses the authority upon which specific support rests (Easton, 1965), showing him as the actor who controls the course of the war, as the supreme commander-in-chief of the Russian army issuing military commands to “bomb Kyiv” (Figure 20). In peace-deal memes, he is shown personally determining the logic of a “dogovornyachok”, usually being demeaning to Ukraine. This representation reaffirms his image as a person who single-handedly determines the course of the war and is reinforced through depictions of Putin along with tanks or submarines to emphasize Russia’s military strength.

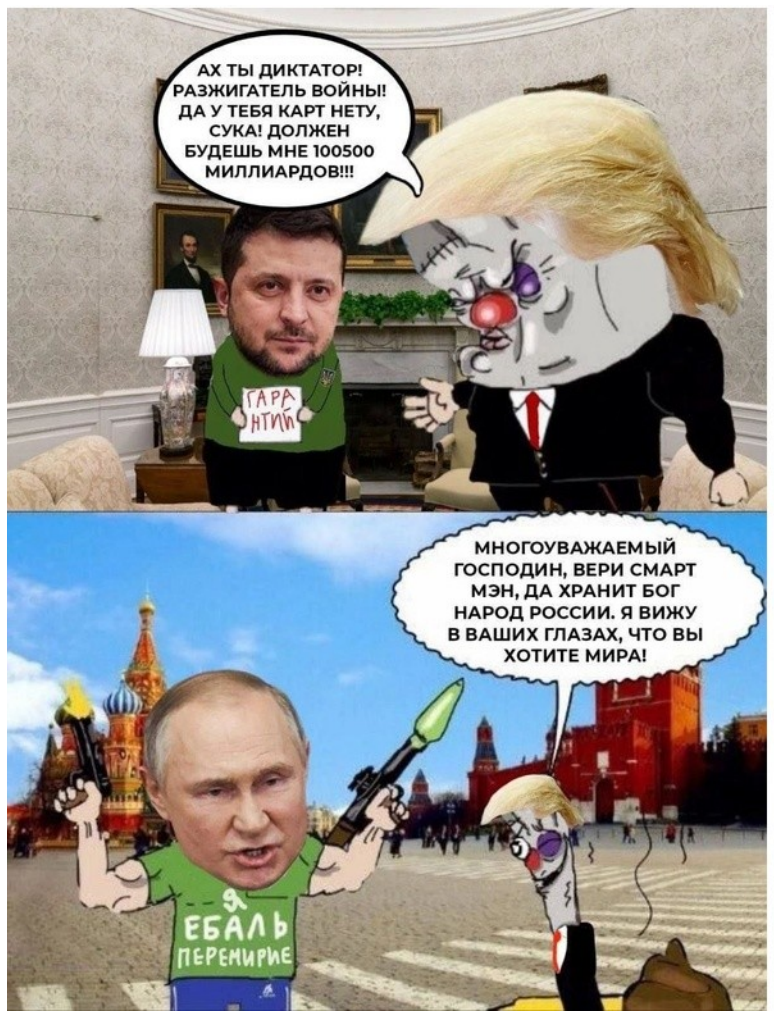

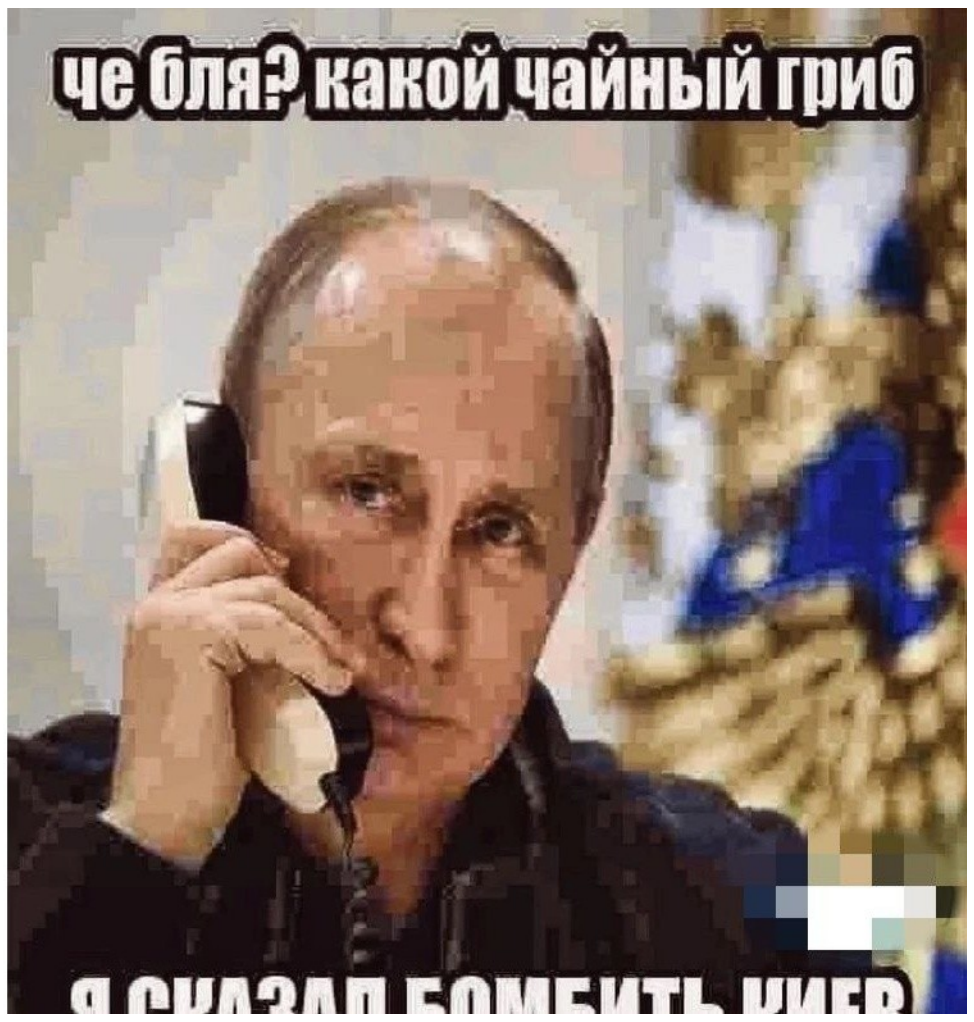

**Figure 21.** Memes depicting Putin as the commander of the war.

*Translation of the texts on the memes:* "You're a dictator! You'll owe me 100,500 billion!" "Very smart man, I see you want peace!"; "What the fuck? I said bomb Kiev."

Similar to Trump, memes show Putin as an internationally recognized political leader, depicting him with many foreign politicians, such as Xi Jinping, Kim Jong Un, or Modi (Figure 20). Particularly prominent are depictions of Putin with them in informal settings, smiling and spending time together. These interpretations emphasize that Putin is a legitimate leader who, following the invasion of Ukraine, continues to act confidently on the global political arena, rejecting Russia's international isolation.

Within this framework, Trump is assigned a special role: he appears as one of the few Western leaders capable of engaging in dialogue with Russia. In some memes, Putin and Trump are depicted in negotiations and casual conversation, with Putin explaining political realities and Trump kindly agreeing with him (Figure 22). Their relationship is additionally portrayed through romanticized imagery and jokes (Figure 10).

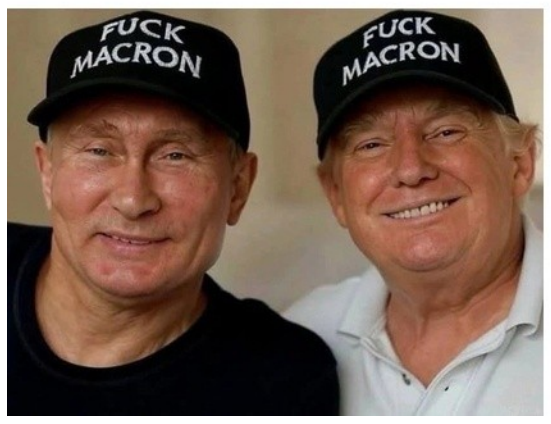

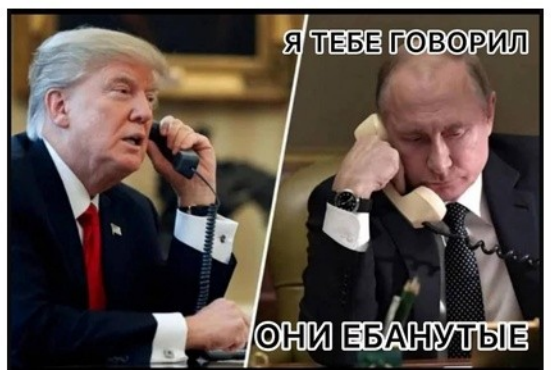

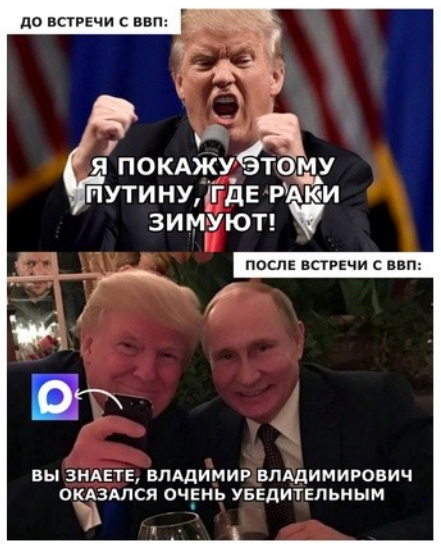

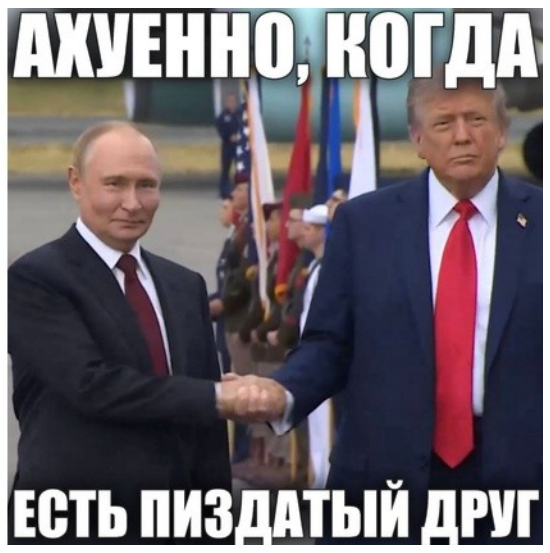

**Figure 22.** Memes depicting Putin and Trump in relaxed settings.

*Translation of the texts on the memes:* "I told you, they're nuts"; "Before meeting Putin: I'll show him! After: he turned out to be very convincing"; "It's great to have an awesome friend."

Some memes depict Putin as a quasi-mythical figure in Warhammer 40k universe. This setting represents an extremely militarized universe in which humanity exists in a permanent state of war. Its central political figure, the God-Emperor of Mankind, is a sacralized ruler legitimizing endless warfare as necessary for survival. Warhammer 40k simultaneously aestheticizes authoritarianism, which makes it particularly popular in far-right discourse. Within this framework, Putin is shown as the God-Emperor, a framing that reinforces his symbolic legitimacy.

Outside military contexts, memes show Putin engaging in outdoor activities (Figure 23). Such representations emphasize his physical capabilities and masculinity. It aligns with broader conservative and nationalist discourse, where political leadership is strongly connected to masculine authority and military strength. A related group of memes stresses the accessibility of Putin and his closeness to ordinary Russians; some of them originate from “Direct Line” broadcasts where Putin answers citizens’ questions or show him at home in pajamas. Such populist representations contrast Putin with other members of Russian (and Western) political elites who are often presented as being too far from the people.

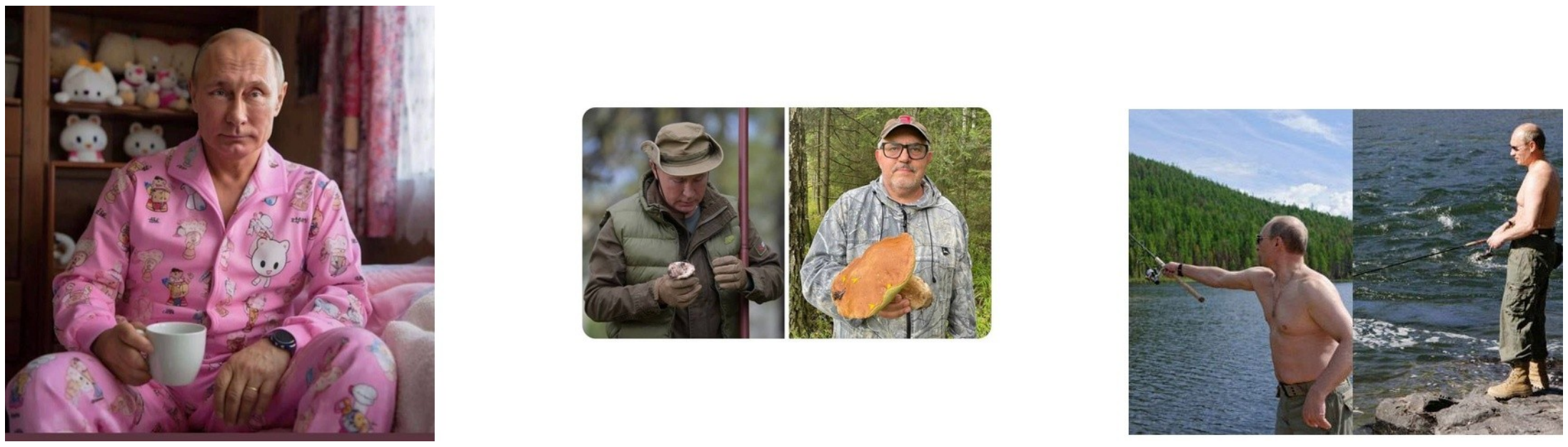

**Figure 23.** Memes depicting Putin outdoors and in everyday settings.

Another cluster serving the same goal of his *humanization* appears in recurring imagery of Putin with animals such as cats, koalas, and capybaras. These depictions usually show him carefully holding animals, presenting Putin as a calm and caring person and complementing the broader image of authority with socially desired human traits.

Classic internet meme aesthetics are also actively incorporated into Putin’s representation. Edits such as “sigma face” or “yassified” versions of Putin (Figure 24) portray him as attractive and confident. Within internet culture, these formats are typically used to amplify charisma and status. In the Russian context, they translate political authority into recognizable forms of online masculinity and prestige.

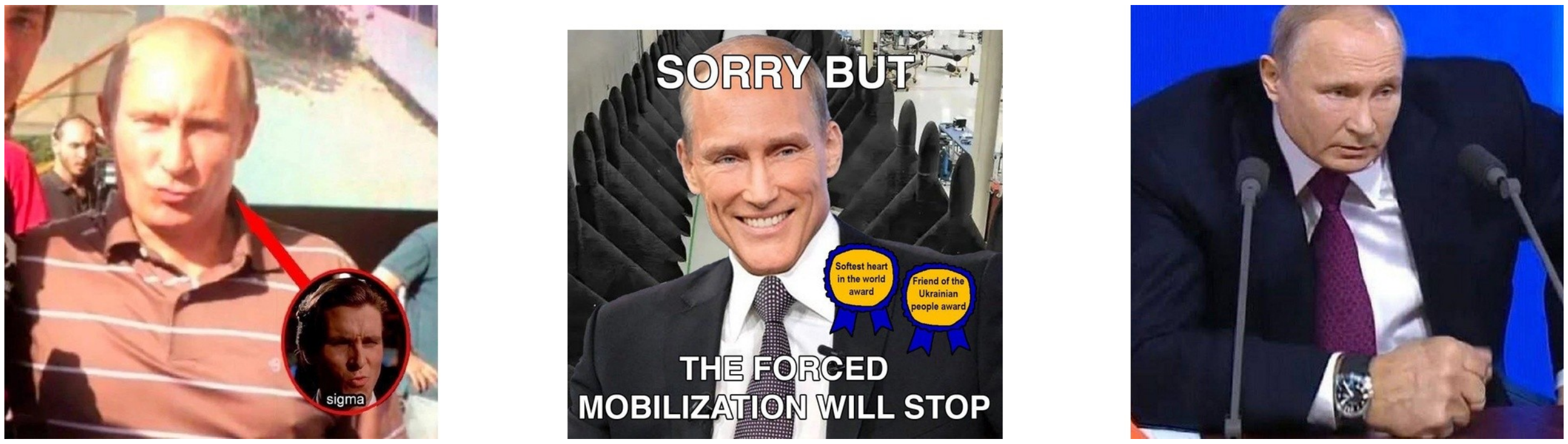

**Figure 24.** Memes depicting Putin in classic internet meme formats.

Some memes criticize Putin, but they rarely substantially challenge his political legitimacy and agency. Criticism includes economic problems or demands to “send sons to war”, but none directly calls out Putin as incompetent. Most memes present war as a necessity, normalizing military sacrifice and countering the dissent potential of jokes related to mobilization or wartime losses. Requests for taxes or military participation are therefore framed less as a form of oppression and more as expected demands issued by a sovereign wartime leader. In many cases, these jokes also function as long-circulating meme formats or “meme magnets” common within Russian internet culture, like a joke about Putin willing to have more taxes.

Other ironic or semi-critical representations include conspiracy theories about body doubles (also a common meme-magnet not only within far-right communities) or portraying Putin as a monkey (Figure 25). Irony here often functions as a form of “safe criticism” that allows limited expression of dissatisfaction without questioning Putin’s right to rule. Such memes remain relatively rare compared to the legitimizing representations dominating the dataset.

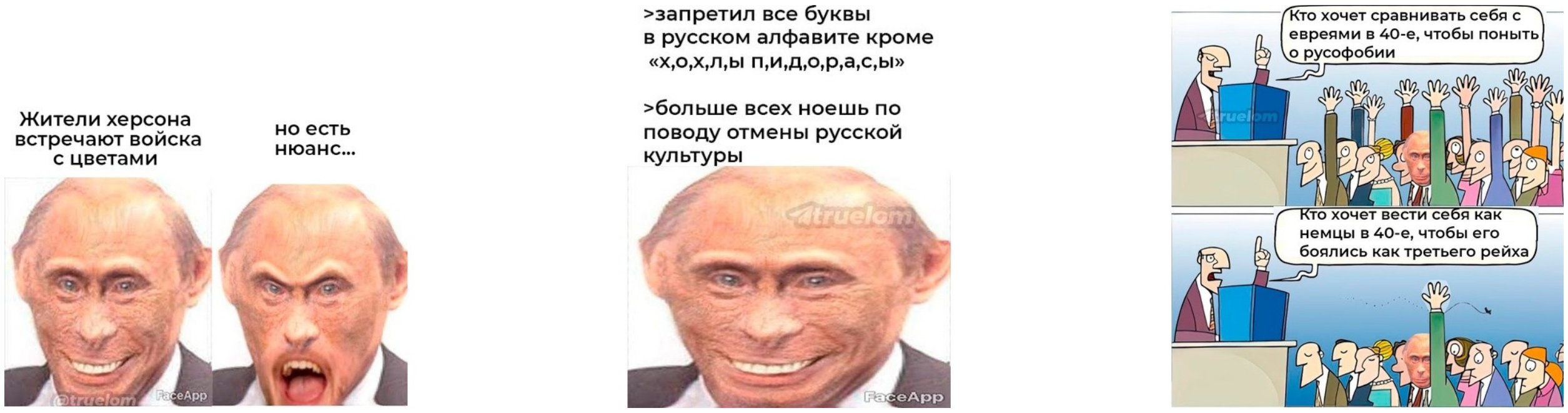

**Figure 25.** Memes depicting Putin ironically or critically.

*Translation of the texts on the memes:* “Residents of Kherson greet the troops with flowers. But there’s a nuance”; “Banned every letter except ‘khokhly are faggots’, whines about the cancellation of Russian culture”; “Who wants to compare themselves to the Jews of the 1940s to whine about Russophobia?”

Overall, Putin’s representation reiterates an image of a globally recognized and authoritative actor. While humor is widely applied, it remains within the confines of this image and does not undermine Putin’s legitimacy, because even in semi-critical images, the integrity of his power and agency is never called into question.

## Conclusion

Consistent with Vaara’s (2014) understanding of legitimation and delegitimation as discursive processes, far-right communities actively participate in shaping representation of political actors through online participatory culture, including memes. Far-right memetic discourse assigns different degrees of political authority and humanity to political actors. Among the examined actors, Putin occupies the position of full legitimization and is represented as a strong, globally respected, and masculine leader. This representation aligns with Russian state propaganda, where political stability and national strength are personalized in Putin’s figure (Malinova, 2022; Szostek, 2017). Trump, despite being a Western political actor, receives a form of legitimization almost comparable to Putin's through memes emphasizing geopolitical influence, anti-liberal sentiment, and support for the Russian president.

Other participants occupy a less privileged position. Medvedev's representation is characterized by controlled delegitimization: although he is frequently mocked, this criticism rarely extends to the regime itself and instead reinforces Putin's superior image. Similarly, Biden is shown as an actor possessing actual power over Zelensky, despite being systematically delegitimized through the discourse focused on incompetence and cognitive decline. Such portrayal of Biden (and, to a degree, Medvedev) aligns with the Kremlin propaganda discourse presenting Western political leadership as ineffective and morally decaying (Gerber & Zavisca, 2016). Zelensky is a target of full delegitimization, with memes referring to him as subordinate, externally controlled, and lacking political agency. It aligns with Kremlin propaganda that presents Ukraine as a Western colony governed by a puppet leadership (Tolz & Hutchings, 2023).

These findings demonstrate how far-right communities participate in the co-production of authoritarian legitimacy by extending and radicalizing state-aligned messages through participatory online culture. It illustrates a broader transformation in information autocracies, where ideological messages are no longer communicated exclusively through centralized top-down mechanisms as in pre-digital authoritarian regimes, but are reshaped and circulated within online communities, including those that are not necessarily directly sponsored by the regime and are more of situational allies. These communities repackage pro-regime narratives through humor, making them more engaging and harder to identify as state propaganda.

In contrast to traditional channels of political communication in informational autocracies, memes operate within more flexible boundaries and often normalize absurd or transgressive forms of expression. Dehumanizing rhetoric about Ukrainians is also present in official discourse (Tolz & Hutchings, 2023), most visibly in Medvedev's statements. Official media, however, rarely use imagery as extreme as the animalistic, objectifying, and antisemitic comparisons, given broader diplomatic considerations, which are less applicable to fringe media actors such as far-right Telegram groups. Similarly, the references to "bromance" between Putin and Trump may be too inappropriate for conservative pro-regime media, but this ironic legitimation strategy is highly expressive when used in memes. Far-right memes therefore constitute a distinct part of the pro-regime information environment, where official ideological narratives are translated into the vernacular language of internet culture (and, in turn, can potentially influence official discourse, where some memetic expressions and tropes can be picked up by more traditional media channels). This process allows political messages to circulate through formats that rely on irony and participatory remixing, while leaving greater room for experimentation and selective criticism of individual political actors.

It is important to note some limitations of the current study and directions for future research. First, we examined memes related to relatively few political actors, whereas far-right memes that we collected also refer to other individuals (e.g., members of Russian political opposition, other pro-Kremlin politicians from Russia and abroad) or groups (e.g., Ukrainians or Russians) when constructing authoritarian legitimacy. Future research can benefit from expanding the range of actors included in the analysis to better understand different (de)legitimation strategies which may be applied to less popular political figures. Second, we focused on the discursive construction of legitimacy, examining how far-right memes represent matters of Russian and international politics, but not the empirical effects of being exposed to these memes. It remains a question for future research to assess whether interaction with memes which, for instance, praise Putin and dehumanize Zelensky, has substantive impact on the perceived legitimacy of these actors and how such impact can be influenced by internet users' individual characteristics.

## Data availability statement

The images analyzed in this study were collected from publicly available Telegram channels (with the list of channels provided in Appendix A). The code, the identifiers of the channels and messages, and the cluster assignments will be deposited in a public repository after acceptance.

## Declaration of interest statement

The authors report that there are no competing interests to declare.

## Funding

No funding was obtained for the reported work.

## Declaration of generative AI use

Image captions were generated with the vision-language model Qwen2-VL-2B-Instruct to support the interpretation and naming of image clusters. During the preparation of the manuscript, the authors used Grammarly (Grammarly, Inc.) to correct spelling and grammar and to improve the language. The authors reviewed and edited all AI-assisted output, kept the original and revised versions of the text, and take full responsibility for the content of the manuscript.

# Appendix A

*Telegram channels included in the dataset*

| | | |
|---|---|---|
| @anime_and_veterans | @istrkalkglk | @szagatin |
| @Baba_s_yajcami_i_memami | @karga4 | @truelom |
| @basedpolitics | @killaknight14228 | @viatka_aristocrat |
| @c_a_x_a_p_H_o_e | @mercOf2k17 | @viatka_memes |
| @Current_memes | @milosevic_online | @voen2ch |
| @DumaiteMemes | @pravie | @war_memez |
| @grpzdc | @proectrazgrom | @warmempolit |
| @gus_vasili | @russian_civilisation_memes | @yasenslez |
| @hahahohly | @russkie_koti | «Правые мемы» https://t.me/+q_wVNO-KoskxZDUy |
| @illegaltie | @sosach103 | |
| @insect_life | @skazochnoebalie | |

# Appendix B

Face recognition results showed high confidence levels: mean confidence scores exceed 98% for all actors, with minimal values remaining above 75%. For downstream analysis, the consolidated dataset was further partitioned into person-specific subsets for five individuals we focused on. To validate the precision, one of the authors manually reviewed a random sample of 100 images for all actors. All manually inspected images were correctly classified, corresponding to an observed precision of 100% for each validation sample.

Image embeddings were generated using the vision-language model Contrastive Language–Image Pretraining (CLIP; Radford et al., 2021), which encodes high-level visual and semantic features. Specifically, the pretrained model "openai/clip-vit-base-patch32" was used via the Hugging Face Transformers library. Images were processed in batches (batch size = 64), converted to RGB format, and encoded using the model's vision encoder. The pooled visual representations were projected into a 512-dimensional embedding space and L2-normalized.

To improve clustering performance, dimensionality reduction was applied using Uniform Manifold Approximation and Projection (UMAP) (Allaoui et al., 2020). The UMAP model was configured with n_neighbors = 10, n_components = 10, min_dist = 0.0, and the cosine distance metric.

Clustering was performed using Hierarchical Density-Based Spatial Clustering of Applications with Noise (HDBSCAN; Allaoui et al., 2020). The model was configured with min_cluster_size = 8, min_samples = 2, the Euclidean distance metric, and cluster_selection_method = "eom". Images assigned to cluster label −1 were excluded from topic-level interpretation. Clustering was performed exclusively on CLIP image embeddings; captions were used only for cluster interpretation and naming and were not included in the clustering procedure.

For each image, automatically generated captions were used as a textual proxy for visual content. Captions were generated using the multimodal vision-language model Qwen2-VL-2B-Instruct (Wang et al., 2024), which produced detailed descriptions of visible elements in each image and provided initial cluster names.